\documentclass[twoside,twocolumn,9pt]{article}
\usepackage{extsizes}
\usepackage[super,sort&compress,comma]{natbib} 
\usepackage[version=3]{mhchem}
\usepackage[left=1.5cm, right=1.5cm, top=1.785cm, bottom=2.0cm]{geometry}
\usepackage{balance}
\usepackage{mathptmx}
\usepackage{sectsty}
\usepackage{graphicx} 
\usepackage{lastpage,soul,babel}
\usepackage[format=plain,justification=justified,singlelinecheck=false,font={stretch=1.125,small,sf},labelfont=bf,labelsep=space]{caption}
\usepackage{float}
\usepackage{fancyhdr}
\usepackage{fnpos}
\addto{\captionsenglish}{%
  
}
\usepackage{array}
\usepackage{droidsans}
\usepackage{charter}
\usepackage[T1]{fontenc}
\usepackage[dvipsnames]{xcolor}
\usepackage{setspace}
\usepackage[compact]{titlesec}
\usepackage{hyperref}
\usepackage{lmodern}
\usepackage{amssymb}

\usepackage{epstopdf}

\definecolor{cream}{RGB}{222,217,201}

\begin{document}

\pagestyle{fancy}
\thispagestyle{plain}
\fancypagestyle{plain}{
\renewcommand{\headrulewidth}{0pt}
}

\makeFNbottom
\makeatletter
\renewcommand\LARGE{\@setfontsize\LARGE{15pt}{17}}
\renewcommand\Large{\@setfontsize\Large{12pt}{14}}
\renewcommand\large{\@setfontsize\large{10pt}{12}}
\renewcommand\footnotesize{\@setfontsize\footnotesize{7pt}{10}}
\makeatother

\renewcommand{\thefootnote}{\fnsymbol{footnote}}
\renewcommand\footnoterule{\vspace*{1pt}%
\color{cream}\hrule width 3.5in height 0.4pt \color{black}\vspace*{5pt}} 
\setcounter{secnumdepth}{5}

\makeatletter 
\renewcommand\@biblabel[1]{#1}            
\renewcommand\@makefntext[1]%
{\noindent\makebox[0pt][r]{\@thefnmark\,}#1}
\makeatother 
\renewcommand{\figurename}{\small{Fig.}~}
\sectionfont{\sffamily\Large}
\subsectionfont{\normalsize}
\subsubsectionfont{\bf}
\setstretch{1.125} 
\setlength{\skip\footins}{0.8cm}
\setlength{\footnotesep}{0.25cm}
\setlength{\jot}{10pt}
\titlespacing*{\section}{0pt}{4pt}{4pt}
\titlespacing*{\subsection}{0pt}{15pt}{1pt}

\fancyfoot{}
\fancyfoot[LO,RE]{\vspace{-7.1pt}\includegraphics[height=9pt]{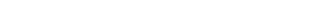}}
\fancyfoot[CO]{\vspace{-7.1pt}\hspace{13.2cm}\includegraphics{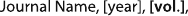}}
\fancyfoot[CE]{\vspace{-7.2pt}\hspace{-14.2cm}\includegraphics{head_foot/RF}}
\fancyfoot[RO]{\footnotesize{\sffamily{1--\pageref{LastPage} ~\textbar  \hspace{2pt}\thepage}}}
\fancyfoot[LE]{\footnotesize{\sffamily{\thepage~\textbar\hspace{3.45cm} 1--\pageref{LastPage}}}}
\fancyhead{}
\renewcommand{\headrulewidth}{0pt} 
\renewcommand{\footrulewidth}{0pt}
\setlength{\arrayrulewidth}{1pt}
\setlength{\columnsep}{6.5mm}
\setlength\bibsep{1pt}
\newcommand\nnfootnote[1]{%
  \begin{NoHyper}
  \renewcommand\thefootnote{}\footnote{#1}%
  \addtocounter{footnote}{-1}%
  \end{NoHyper}
}
\makeatletter 
\newlength{\figrulesep} 
\setlength{\figrulesep}{0.5\textfloatsep} 

\newcommand{\topfigrule}{\vspace*{-1pt}%
\noindent{\color{cream}\rule[-\figrulesep]{\columnwidth}{1.5pt}} }

\newcommand{\botfigrule}{\vspace*{-2pt}%
\noindent{\color{cream}\rule[\figrulesep]{\columnwidth}{1.5pt}} }

\newcommand{\dblfigrule}{\vspace*{-1pt}%
\noindent{\color{cream}\rule[-\figrulesep]{\textwidth}{1.5pt}} }

\makeatother

\twocolumn[
  \begin{@twocolumnfalse}
{\includegraphics[height=30pt]{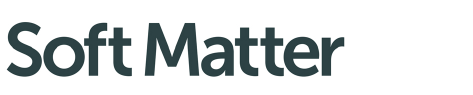}\hfill\raisebox{0pt}[0pt][0pt]{\includegraphics[height=55pt]{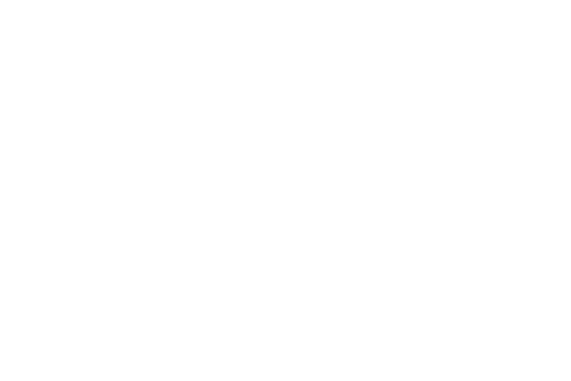}}\\[1ex]
\includegraphics[width=18.5cm]{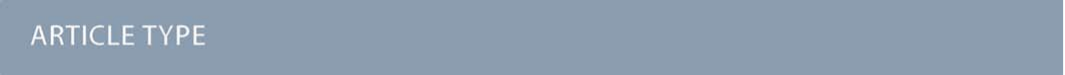}}\par
\vspace{1em}
\sffamily
\begin{tabular}{m{4.5cm} p{13.5cm} }

\includegraphics{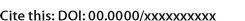} & \noindent\LARGE{\textbf{Droplet Breakup Against an Isolated Obstacle}} \\
\vspace{0.3cm} & \vspace{0.3cm} \\

 & \noindent\large{David J. Meer,\textit{$^{a,*}$} Shivnag Sista,\textit{$^{b,*}$} Mark D. Shattuck,\textit{$^{c}$} Corey S. O'Hern,\textit{$^{bdef}$} and Eric R. Weeks\textit{$^{a}$}} \\

\includegraphics{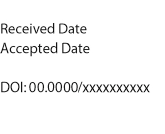} & \noindent
\normalsize{We describe combined experiments and simulations of single droplet breakup during flow-driven interactions with a circular obstacle in a quasi-two-dimensional microfluidic chamber. Due to a lack of in-plane confinement, the droplets can also slip past the obstacle without breaking.  Droplets are more likely to break when they have a higher flow velocity, larger size (relative to the obstacle radius $R$), smaller surface tension, and for head-on collisions with the obstacle. We also observe that droplet-obstacle collisions are more likely to result in breakup when the height of the sample chamber is increased.
We define a nondimensional breakup number ${\rm Bk} \sim {\rm Ca}$ that accounts for changes in the likelihood of droplet break up with variations in these parameters, where ${\rm Ca}$ is the Capillary number.  As ${\rm Bk}$ increases, we find in both experiments and discrete element method (DEM) simulations of the deformable particle model that the behavior changes from droplets never breaking (${\rm Bk} \ll 1$) to always breaking for ${\rm Bk} \gg 1$, with a rapid change in the probability of droplet breakup near ${\rm Bk} = 1$. We also find that Bk $\sim S^{4/3}$, where $S$ characterizes the symmetry of the collision, which implies that the minimum symmetry required for breakup is controlled by a characteristic distance $h \sim R$.} 

\end{tabular}

 \end{@twocolumnfalse} \vspace{0.6cm}

]
\renewcommand*\rmdefault{bch}\normalfont\upshape
\rmfamily
\section*{}
\vspace{-1cm}

\begin{NoHyper}
\footnotetext{\textit{$^*$~Contributed equally and share first authorship.}}
\footnotetext{\textit{$^{a}$~Department of Physics, Emory University,  Atlanta, GA 30322, USA. Email: dmeer@emory.edu}}
\footnotetext{\textit{$^{b}$~Department of Physics, Yale University, New Haven, Connecticut, 06520, USA}}
\footnotetext{\textit{$^{c}$~Benjamin Levich Institute and Physics Department, The City College of New York, New York, New York 10031, USA}}
\footnotetext{\textit{$^{d}$~Department of Mechanical Engineering, Yale University, New Haven, Connecticut, 06520, USA}}
\footnotetext{\textit{$^{e}$~Department of Applied Physics, Yale University, New Haven, Connecticut, 06520, USA}}
\footnotetext{\textit{$^{f}$~Department of Materials Science, Yale University, New Haven, Connecticut, 06520, USA}}
\end{NoHyper}



\section{Introduction}

Droplet formation is key to mixing two immiscible liquids to form an emulsion,\cite{bibette99} spread of some transmissible diseases via airborne droplets,\cite{poon20} and inkjet printing.\cite{tabeling10}  Furthermore, microfluidic devices are used to form droplets for lab-on-a-chip applications.\cite{selimovic10}  Making droplets often involves starting with two liquids, adding energy by shaking, stirring, or otherwise flowing the two liquids, and thus mixing the fluids into large droplets of one fluid mixed into the other.  Previous work has studied how  fluid flow patterns at the individual droplet scale cause large droplets to break into smaller droplets.  The simplest case is an isolated droplet in a shear flow.\cite{Janssen1994}  The surface tension of the droplet tries to minimize its surface area, and thus acts to maintain a spherical shape.  Competing with surface tension, viscous stresses caused by the fluid shear flow try to stretch the droplet.  For sufficiently fast flows, these viscous forces make the droplets deform or even tear themselves apart into smaller droplets.  The capillary number is a dimensionless parameter that measures the ratio of viscous forces to surface tension forces
\begin{equation}
    \mathrm{Ca}=\frac{\mu v}{\gamma},
    \label{eqn:capillary}
\end{equation}
where $\gamma$ is the surface tension and the viscous forces are quantified by the continuous phase dynamic viscosity $\mu$ and a characteristic flow velocity $v$.  For $\mathrm{Ca} > \mathrm{Ca}_c$, droplet breakup occurs, where the critical value $\mathrm{Ca}_c$ depends on the specific geometry of the fluid flow.\cite{Janssen1994,link_geometrically_2004,fu_dynamics_2011}  Prior work has studied droplet breakup in relatively simple microfluidic geometries, for example, in T-junctions,\cite{link_geometrically_2004,cheng_prediction_2018,fu_dynamics_2011,wankawala_experimental_2023,wankawala_enhanced_nodate,tiribocchi_shapes_2023, Wankawala2025EPL} constrictions where droplets drip from a nozzle,\cite{Utada2007} and narrow channels with an obstacle in the middle where the droplet wraps around both sides of the obstacle and then breaks.\cite{protiere_droplet_2010} In these prior experiments, at large Ca viscous effects dominate causing increased droplet breakup. At small Ca, the droplets can deform, but they do not breakup.  

Droplet breakup is less understood in more complex geometries, such as porous media. The flow of two immiscible fluids through porous media is important for numerous applications, such as petroleum extraction, \cite{Betancur_2024,Durlofsky2005} pharmaceutical manufacturing,\cite{Saitoh2008,Browne2024} and agricultural and food production.\cite{schroen_droplet_2021}  Droplets moving through porous media are also crucial for understanding the flow of groundwater pollutants, such as PFAS.\cite{chen_porescale_2023,Liu2021}  One key feature of porous media is that droplets can be found in channels larger than their diameter, allowing them to assume complex shapes not observed in more confined geometries.

Many previous studies have considered flows through fully wetted porous media (e.g. oil fills the pore space) that is invaded by an immiscible fluid (water).\cite{datta13prl,datta13,datta14,datta14b}  The effect of surface tension is enhanced if one of the fluids forms droplets, thus greatly increasing the interfacial area between the two fluids. A stream of droplets moves differently through a porous medium compared to invasion of a continuous phase fluid into a porous medium,\cite{parrenin_effect_2024} mainly due to the increase in interfacial area. Previous studies of droplets flowing through porous media have in many cases not considered deformable droplets and droplets that can break up.\cite{izaguirre_emergence_2024,Benet2017,benet_mechanical_2018} 

In this article, we seek to understand droplet flow through a simplified porous medium, i.e., a single droplet interacting with a single obstacle, which is droplet flow through a porous medium in the limit of small droplet and obstacle densities. (See Supplementary Video 1.) In  Fig.~\ref{fig:intro}, we show that the droplets can either move around the obstacle or wrap around it and break into two smaller droplets due to the flow. Several parameters determine whether the droplet breaks up or not. The first parameter influencing breakup is Ca. Faster flows have larger viscous forces that push the droplet against the obstacle, and surface tension prevents the droplet from deforming.  Second, larger droplets are easier to deform and break, so the droplet size is an important parameter. A third parameter is the symmetry of the collision between the droplet and obstacle.  Head-on collisions of the droplets with the obstacle lead to a higher probability of breakup; for example, compare Fig.~\ref{fig:intro}(a,b) with (e,f).  The symmetry of the collision was not relevant in prior experimental studies.~\cite{protiere_droplet_2010,Stone2013,cheng_prediction_2018,fu_dynamics_2011} Droplets can also slide around the obstacle, requiring only small droplet deformations as shown in Fig.~\ref{fig:intro}(e,f).

\begin{figure}
    \centering
    \includegraphics[width=0.665\linewidth]{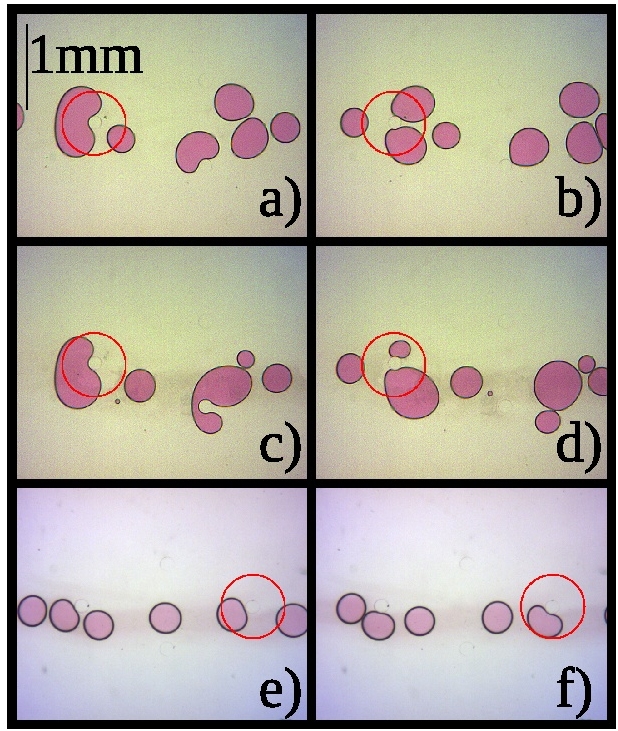}
    \includegraphics[width=0.325\linewidth]{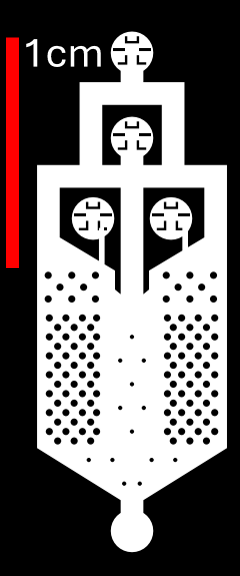}
    
    \caption{(Left) Images of droplets moving from left to right through the array of obstacles with collisions highlighted by red circles in each image pair. (a-b) A head-on collision (with symmetry $S=1.0$) causes a large droplet with velocity $v=1.8$~mm/s to break up; (c-d) An asymmetric collision with $S=0.5$ between a large droplet with $v=1.4$~mm/s causes droplet break up; and (e-f) An asymmetric collision with $S=0.5$ between a small droplet with $v=1.1$~mm/s does not lead to break up. The time interval $\Delta t$ between the left and right images in each row is determined by the velocity $v$ so that the distance traveled is fixed, $v \Delta t=400~\mu$m. (Right) Microfluidics design of the sample chamber, which is $\approx 22$~mm long. Droplets form at the middle top region and exit through the central channel into the wider region below.  In the wider region, droplets collide with small obstacles before exiting the chamber at the bottom outlet. }
    \label{fig:intro}
\end{figure}

We also develop mesoscale simulations to model droplet shape evolution and breakup due to stresses arising from interactions between the droplets and the obstacles and continuous phase fluid. For the simulations, we will employ the deformable particle model with surface tension~\cite{cheng24} supplemented by a geometric criterion for the onset of droplet breakup. An advantage of the deformable particle model is that it includes only a small number of physics-informed parameters that can be calibrated to the experimental results. The deformable particle simulations will allow us to map the regions of parameter space, such as the continuous phase fluid viscosity, droplet surface tension, droplet-to-obstacle size ratio, and collision symmetry, where droplet break up does and does not occur. Another advantage of the deformable particle simulations is that they enable exploration of regions of parameter space that are difficult to access experimentally. In particular, the simulations allow independent variation of the surface tension and continuous phase fluid viscosity, so that we can isolate and quantify their individual effects on droplet deformation and breakup. In contrast, in the experiments, the propensity for droplet breakup is tuned indirectly through changes in the flow conditions, effectively varying the capillary number. 

The remainder of this article is organized as follows. In Section 2, we describe the experimental setup and method to generate, control, track, and quantify droplet-obstacle collisions. In Section 3, we describe mesoscale simulations of fluid-flow driven droplet-obstacle interactions using the deformable particle model with a geometric criterion for droplet breakup. In Section 4, we quantify when droplet break up occurs in the experiments and simulations by defining a nondimensional ``breakup number'' Bk, which is Ca multiplied by the ratio of the area of the droplet to that of the obstacle, and other geometric factors. Thus, droplet breakup is more likely at higher Ca, for droplets that are larger relative to the obstacle, and droplets that incur head-on collisions with the obstacle. Section 5 concludes with a summary of the results and promising future research directions.  We include three appendices that provide details of the microfluidics device design (Appendix A), validation of the continuous phase fluid model used in the simulations (Appendix B), and independent variation of the surface tension and viscosity in the simulations to determine their effects on droplet breakup (Appendix C). 

\section{Experimental Methods}
\label{sec:exptmethods}

The experimental flow cell, as shown in Fig.~\ref{fig:intro}, consists of droplets, a continuous phase fluid flow, and obstacle pillars. The droplets are composed of water, a rhodamine dye added to saturation, and 1\% tween-20 by mass, well above the critical micelle concentration. We did not observe depletion forces, since the surfactant is in the droplet phase and there was a constant flow of fresh silicon oil into the cell. The results were replicated using a grocery-store food coloring instead of rhodamine, suggesting that the properties of the specific dye are unimportant.  The surfactant is included to provide a barrier to coalescence, although occasionally coalescence events are observed and these events are excluded from the data analysis.  The continuous phase fluid phase is one of two silicon oils with kinematic viscosity $\nu_{\rm oil} = 50$~cSt or 100~cSt. Both oils have density $\rho_{\rm oil}=960$~kg/m$^3$. These values yield a dynamic viscosity of $\mu=\nu \rho = 480$ or 960~Pa$\cdot$s depending on which silicon oil is used. The surface tension between the continuous phase fluid and droplets is $\gamma \approx 20$~mN$/$m.  The silicon oil is injected at a flow rate of $20$-$60~\mu$L/min, driving the droplets with a velocity in the range of $v = 0.5$ to $2$~mm/s. The water is injected at a flow rate in the range of $15$ to $120$~$\mu$L/hr, which creates droplets using the pinch-off effect.\cite{Utada2007}  The droplet diameters are in the range $D_0 \approx 100$ to $600~\mu$m, and the obstacle radii are in the range $R = 60$ to $120$~$\mu$m. Based on these values, the Reynolds number, defined as ${\rm Re}= \rho_{\rm oil} v R / \eta_{\rm oil}$ is $\leq O(10^{-2})$ for all experiments, indicating that inertial effects are negligible.

The microfluidic chamber is made from polydimethylsiloxane (PDMS) created by pouring a degassed mixture of unsolidified polymer and curing agent (7.5:1 ratio by mass) onto a silicon wafer with the desired pattern, shown in Fig.~\ref{fig:intro}, etched into it. This etching is performed using photolithography on an SU-8 surface, with a photomask ordered from ARTNET Pro, Inc. A profilometer measured the depth of the etching on the silicon wafer to be $z=85~\mu$m $\pm5~\mu$m, which sets the sample chamber thickness $z$. Another sample chamber with $z=45~\mu$m $\pm5~\mu$m is used for a subset of experiments to test the influence of $z$ on the results. The PDMS is then allowed to solidify on this etching, either over a weekend or overnight with 70$^{\circ}$C heating.  After the PDMS chambers are prepared and cut from the wafer, they are bonded to a microscope slide covered in PDMS using oxygen plasma cleaning, which allows it to serve as the ``floor" of the chamber.\cite{barnes_plasma_2021}

To avoid the need for 3D imaging, we created quasi-2D water-in-oil droplets with volume $V$ that satisfies $\sqrt[3]{V}\gg z$. Thus, the droplets are ``pancake'' shaped with small out-of-plane curvature, which is observable in Fig.~\ref{fig:intro} as a dark ``border ring."  The outline of the droplets aids in detecting when one region contains a single concave droplet versus two convex droplets. 

In untreated sample chambers, we observe that water droplets can stick to the PDMS chambers.  To prevent adhesive forces, we coat the chambers with Aquapel which increases the hydrophobicity of the surfaces.  The process works best when Aquapel flows through and is heated to 70$^{\circ}$C \cite{nawar_parallelizable_2020} for at least $20$ minutes. We also find that fresh Aquapel creates a less hydrophobic surface than Aquapel stored in a degassed syringe for at least $24$ hours.  In addition, Aquapel spoils, visibly changes color, and loses its hydrophobicity after $\approx$ 2 weeks. Thus, we always use Aquapel within a few days after it arrives at the laboratory, but after storing overnight in the syringe.

The microfluidic chamber experiments are recorded using a LEICA DMIRB microscope at 60 frames per second with a ThorLabs DCC1645C - USB 2.0 CMOS camera at $640 \times 512$ resolution. We used a $1.6\times$ objective (0.05 Numerical Aperture air lens) and a $1.5 \times$ zoom, resulting in videos with a scale of $5.28~\mu$m/pixel. These videos are processed by separating the images  into three regions: the continuous phase which is ignored, the pink centers of the droplets, and the black border of each droplet, which is assigned to the pink region that the border encircles.  We can detect when two or more droplets are in contact, and we reject droplets that are in contact with other droplets during collisions with an obstacle. After segmentation and particle tracking,\cite{crocker96} we have data on the droplet area, velocity, position relative to obstacles, whether the droplet broke, and if so, the sizes of the daughter droplets after the collision for 5,056 droplet-obstacle collisions.

The droplet-obstacle collisions are obtained under several experimental conditions.  The standard parameters are the following: obstacle radius $R=85$~$\mu$m, continuous phase dynamic viscosity $\mu_{\rm oil}=480$~Pa$\cdot$s, and sample chamber thickness $z=85$~$\mu$m.  In addition to the standard set of parameters, we also vary each parameter one at a time, investigating $\mu_{\rm oil}$=960~Pa$\cdot$s by changing the silicon oil, $R$=60~$\mu$m and 120~$\mu$m, and $z$=45~$\mu$m.  Each experimental movie (Supplementary Video) contains multiple instances of droplet-obstacle collisions both with and without break up, depending on the experimental conditions. The velocity of droplets is modified by changing the continuous phase fluid pump rate, and the size of droplets is modified by controlling the relative flow rates between the continuous phase and droplet fluid \cite{Utada2007}. The angle of collision (later defined as the symmetry parameter in Sec.~\ref{results}) is spontaneously varied by droplets within each experiment as they move through the arrays since neighboring droplets slightly modify each others' flow paths.

\section{Simulation Methods}

\subsection{Deformable Particle Model}
We performed simulations of a single droplet colliding with a single obstacle using the deformable particle model, which can accurately model large deformations of capillary droplets flowing through confined geometries.\cite{cheng24}  In two-dimensions, the droplet is defined as a deformable polygon with $N_v$ vertices, whose positions and velocities are the degrees of freedom of the system. (See Fig.~\ref{fig:DP_model}.) The mass of the droplet is uniformly distributed among the vertices, and the motion of the vertices is determined by the droplet shape-energy function:
\begin{equation}
    U_{s} = \frac{k_a}{2}\left(A-A_{\rm eq}\right)^2  + \frac{k_lN_v}{2}\sum_{i=1}^{N_v}\left(l_i-l_{\rm eq}\right)^2 + U_{\gamma} .
    \label{shape}
\end{equation}

The first term in eqn~(\ref{shape}) imposes a harmonic energy penalty for changes in the droplet area $A$ from the equilibrium value $A_{\rm eq}$ and $k_a$ controls the fluctuations in the droplet area. This term represents the analog of the bulk modulus of the droplet in 2D. The second term in the shape-energy function imposes a harmonic energy penalty for deviations in the separations $l_{i}$ between adjacent vertices $i$ and $i+1$ from the equilibrium length $l_{\rm eq}$ (which is also the diameter of each of the vertices) and $k_l$ controls fluctuations in $l_i$. This term ensures that the vertices are evenly distributed on the droplet surface, preventing them from clumping when the droplet interacts with the obstacle. The factor of $N_v$ in the numerator of the second term of eqn~(\ref{shape}) makes $U_s$ independent of $N_v$. The third term is the energy arising from line tension. We observe in the experiments that the droplet and the obstacle are coated by a thin layer of the continuous phase fluid (i.e. oil) and hence
\begin{equation}
    U_{\gamma} = \gamma_{2D} P = \gamma_{2D} \sum_{i=1}^{N_v} l_i,
\end{equation}
where $\gamma_{2D}\sim \gamma z$ is the line tension corresponding to the oil–water interface, and $P$ is the droplet perimeter.

To prevent overlap between the droplet and the obstacle, we assume that the droplet interacts with the obstacle via pairwise, purely repulsive spring interactions between the obstacle and each of the droplet vertices:
\begin{equation}
    U_{w} =\sum_{i=1}^{N_v}\frac{\epsilon_{w}}{2} (1-2d_i/l_{\rm eq})^2\Theta(1-2d_i/l_{\rm eq}),
\end{equation}
\noindent where $d_i$ is the distance between the center of the vertex $i$ and the surface of the obstacle and $\epsilon_w$ sets the scale of the repulsive interactions. The Heaviside step function $\Theta(\cdot)$ ensures that the force is non-zero only when vertex $i$ overlaps with the obstacle. The total potential energy $U$ of the droplet is given by the sum of the shape-energy function and droplet-wall interaction energy:
\begin{equation}
    U = U_{w} + U_{s}.
\end{equation}

\subsection{Modeling the Effect of the Continuous Phase Fluid} \label{sec: Fluid flow}

To mimic the experiments, the fluid flow in the simulations is pressure-driven. We neglect the effect of the droplet on the continuous phase fluid profile, but we include the drag force on each droplet vertex $i$ from the fluid flow,
\begin{equation}
    {\vec F}^i_{f} = -\frac{\mu D_0}{N_v}({\vec v}_i - {\vec v}_f) \label{eq: drag_force},
\end{equation}
where ${\vec v}_i$ is the velocity of vertex $i$, ${\vec v}^i_f$ is the velocity of the fluid at vertex $i$, $\mu$ is the fluid viscosity, and $D_0 =\sqrt{4A_{\mathrm{eq}}/\pi}$ is the diameter of the undeformed droplet. In eqn~(\ref{eq: drag_force}), the factor of $1/N_v$ ensures that the drag force on the droplet is independent of the number of vertices. To model the flow field, we use

\begin{align}
v_{f,r}(r,\theta)
&= v_{f\infty}\left[
\ln\!\left(\frac{r}{R}\right)
-\frac{1}{2}
+\frac{1}{2}\frac{R^2}{r^2}
\right]\cos\theta, \label{v_r fluid} \\
v_{f,\theta}(r,\theta)
&= -\,v_{f\infty}\left[
\ln\!\left(\frac{r}{R}\right)
+\frac{1}{2}
-\frac{1}{2}\frac{R^2}{r^2}
\right]\sin\theta \label{v_theta fluid} ,
\end{align}

\noindent which enforces no-slip boundary conditions on the surface of the obstacle for the radial $v_{f,r}$ and angular $v_{f,\theta}$ components of ${\vec v}_f$. $v_{f \infty}$ is the velocity of the fluid far from the obstacle. The coordinate system is defined so that the origin is at the center of the obstacle, the horizontal axis is aligned with the fluid flow, $r$ is the distance from the origin, and $\theta$ is the angle relative to the horizontal axis. Eqns (\ref{v_r fluid}) and (\ref{v_theta fluid}) represent Stokes' solution for creeping flow around a circular obstacle, which provide an accurate description of the fluid flow close to the obstacle. We observe that when the droplets are much smaller than the obstacle, this choice for the flow field yields an accurate trajectory for the droplet around the obstacle, since the droplets effectively act as passive tracers, advected by the continuous-phase flow with no appreciable disturbance to the flow field. For droplets that are much larger than the obstacle, the droplet can distort the flow field. Nevertheless, we find that the deformation of large droplets is insensitive to the form of the flow field in these studies that are in the low Reynolds number regime, provided that the no-slip boundary condition at the obstacle is satisfied. (See Appendix B.)

\subsection{Mesoscale Modeling of Droplet Breakup}
\label{mesoscale}

When a droplet in a shallow microfluidic channel encounters an obstacle, the confinement forces it to deform around the obstacle and produce a neck that thins as the in-plane deformation increases. The neck thickness decreases through a combination of viscous drainage and capillary pressure gradients set by the channel height. Once the neck reaches a critical thickness at which capillary stresses can no longer sustain a connected interface across the confined gap, small perturbations at the interface grow and the neck ruptures, resulting in break up of the droplet. In our 2D deformable particle simulations, we use a simple, geometric criterion to determine when a droplet breaks up. 

\begin{figure}
    \centering
    \includegraphics[width=0.95\linewidth]{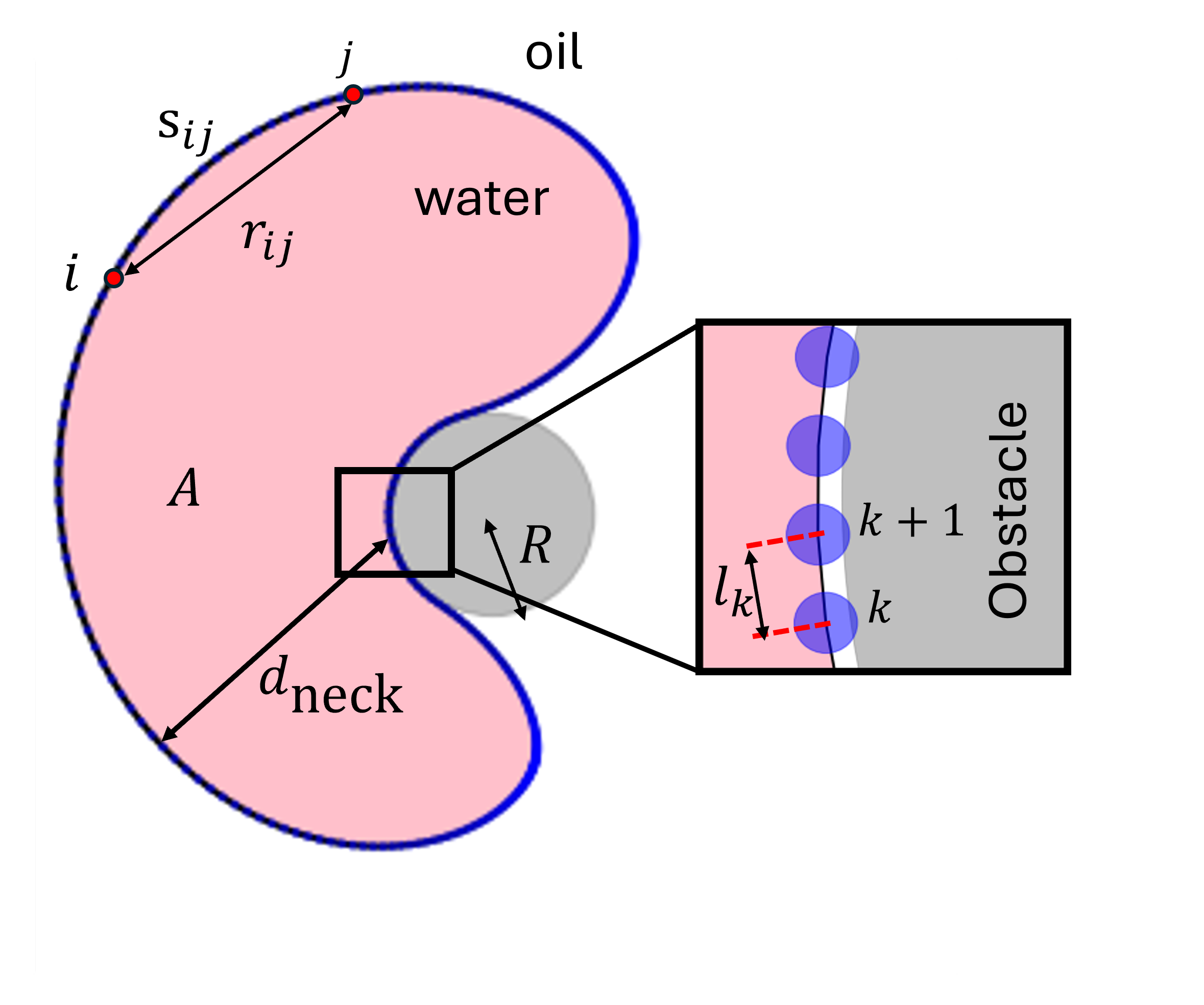}
    \caption{Schematic of a droplet (shaded pink) with area $A$ interacting with an obstacle (shaded gray) with radius $R$. The inset highlights the vertices that define the droplet surface in the deformable particle model. We also define vertex center-to-center distance $r_{ij}$ and the arc length $s_{ij}$ between vertices $i$ and $j$. The droplet neck thickness, $d_{\rm neck}$, is identified using the method described in Sec.~\ref{mesoscale}.}
    \label{fig:DP_model}
\end{figure}

We define the neck thickness $d_{\rm neck}$ as the smallest of the center-to-center distances $r_{ij}$ between every pair of vertices $i$ and $j$ subject to the constraint that the length $s_{ij}$ of the shortest arc joining them (as measured along the perimeter of the droplet) is larger than a threshold $d_{\text{min-sep}}=0.28D_0$. (See Fig.~\ref{fig:DP_model}.) Minimizing $r_{ij}$ subject to this constraint on $s_{ij}$ prevents unphysical breakup events. Specifically, in the limit $d_{\text{min-sep}} \to 0$, daughter droplets form with an unrealistically small number of vertices. Conversely, as $d_{\text{min-sep}} \to D_0$, it becomes increasingly difficult to identify a vertex pair whose shortest connecting arc length exceeds the threshold. Consequently, $d_{\text{min-sep}}$ must fall within a small range. If the neck thickness falls below a smaller threshold $d_{\rm neck} < d_{\rm min-neck}$ at any point during the simulation, we break the droplet into two daughter droplets along the line defining the neck. The value of $d_{\rm min-neck}$ can be calibrated to experimental results for the likelihood of droplet breakup. We find that setting $d_{\rm min-neck}=0.17D_0$ results in the best match to the experimental results.

\subsection{Definitions of the Model Parameters}

In the simulations, we use $D_0$ as the characteristic length scale and $t_0 = D_0/v_{f \infty}$ as the characteristic time scale. Using these along with the droplet mass $M$, we define the dimensionless viscosity $\widetilde{\mu} = \mu t_0D_0/M$, line tension $\widetilde{\gamma}_{2D} = \gamma_{2D} t_0/(Mv_{f \infty})$, edge-length spring constant $\widetilde{k}_l = k_lD_0/\gamma_{2D}$, and area spring constant $\widetilde{k}_a = k_aD_0^4/(Mv_{f \infty}^2)$. We impose fluid incompressibility of the droplet by setting $\widetilde{k}_a > 10^4$ and fix $\widetilde{k}_l/\widetilde{\gamma}_{2D} < 0.05$ so that the line tension energy dominates the perimeter spring energy. 
In a Hele–Shaw geometry \cite{10.1063/1.4952398} such as that used in the experiments, the deformation of a pancake-shaped droplet is governed by a balance between viscous pressure variations induced by confinement and the restoring capillary pressure associated with interfacial curvature. Because the flow is pressure driven, the dominant viscous forcing acting on the droplet originates from lubrication pressure within the thin wetting film that separates the droplet interface from the confining plates. This lubrication pressure $p$ satisfies \cite{Schlichting1968BoundaryLayer}
\begin{equation}
\frac{\partial p}{\partial x} = \mu \frac{\partial^2 v}{\partial y^2},
\end{equation}
where $x$ denotes the flow-direction and $y$ is the direction normal to the flow. If the characteristic thickness of the lubrication layer is denoted by $\delta$, a simple scaling argument gives us
\begin{equation}
\frac{\Delta p_{\mathrm{visc}}}{D_0} \sim \mu \frac{v}{\delta^2},
\end{equation}
where $v$ is the droplet-velocity.
The restoring pressure scale associated with surface tension is set by the inverse radius of curvature of the interface,
\begin{equation}
\Delta p_{\mathrm{cap}} \sim \frac{\gamma}{D_0}.
\end{equation}
The degree of droplet deformation is therefore controlled by the ratio of these two pressure scales,
\begin{equation}
\mathrm{Ca}_{\mathrm{eff}}
\equiv
\frac{\Delta p_{\mathrm{visc}}}{\Delta p_{\mathrm{cap}}}
\sim
\left( \frac{\mu v}{\gamma} \right)
\left( \frac{D_0}{\delta} \right)^2
=
\mathrm{Ca}
\left( \frac{D_0}{\delta} \right)^2,
\label{eq:Ca_eff}
\end{equation}
where $\mathrm{Ca}$ is the experimentally measured capillary number.
Classical lubrication theory predicts that, for a Hele–Shaw gap of height $z$, the lubrication film thickness scales as $\delta \sim z \mathrm{Ca}^{2/3}$.\cite{Shukla_Kofman_Balestra_Zhu_Gallaire_2019, Bretherton_1961, 10.1093/qjmam/12.3.265} In the present experiments, $\mathrm{Ca}=O(10^{-3})$, which implies
\begin{equation}
\frac{\delta}{z} = O(10^{-2}).
\end{equation}
Since $D_0$ and $z$ are of the same order-of-magnitude in our setup, it follows that
\begin{equation}
\frac{\mathrm{Ca}}{\mathrm{Ca}_{\mathrm{eff}}}
=
\left( \frac{\delta}{D_0} \right)^2
=
O(10^{-4}),
\label{eq:approx_conversion}
\end{equation}
This result is consistent with the strong amplification of the viscous stresses induced by confinement.
In the simulations, we define a capillary number $\mathrm{Ca}_{\mathrm{sim}} \equiv \widetilde{\mu}v_{f\infty}/\widetilde{\gamma}_{2D} \sim \left(\mu v_{f\infty}/\gamma\right)(D_0/z)$. Here, we used the fact that $\gamma_{2D} \sim \gamma z$, with the additional length scale $D_0$ arising because the line tension has different physical dimensions than the surface tension. The viscous force in our simulation acts on each vertex and therefore, the viscosity coefficient being used here corresponds to the local viscous forcing that gives rise to the amplification of viscous stresses induced by confinement between two plates. Thus, we can write:
\begin{equation}
    {\rm{Ca}_{\rm{sim}}} \sim {\rm{Ca}_{\rm{eff}}}\frac{D_0}{z} \sim {\rm {Ca_{eff}}},
\end{equation}
\noindent since $z \sim D_0$ are the same order of magnitude. Thus, the capillary number in the 2D simulations corresponds to the effective capillary number $\rm{Ca}_{\rm{eff}}$ (not Ca), and a direct comparison between experiments and simulations requires the conversion \begin{equation}
\mathrm{Ca}
=\mathrm{Ca}_{\mathrm{sim}}\left( \frac{\delta}{D_0} \right)^2. \label{"Ca_sim to Ca"}
\end{equation}
Determining the precise value of the conversion factor $\left(\delta/D_0\right)^2$ is nontrivial, as it depends on details of the lubrication film that are not directly accessible in the experiments. We therefore treat this value as a conversion factor that relates the simulation capillary number $\rm{Ca}_{\rm{sim}}$ and the experimentally measured $\rm{Ca}$. Its value is determined by requiring that the onset of droplet breakup occurs at the same value of the capillary number in the simulations and experiments, with all other parameters held fixed. This procedure gives
\begin{equation}
\label{conv}
    \mathrm{Ca} = 9.8 \times 10^{-5}\,\mathrm{Ca}_{\mathrm{sim}} ,
\end{equation}
which is consistent with eqn~(\ref{eq:approx_conversion}). Henceforth, when discussing the simulation results, we report $\mathrm{Ca}$ from eqn~(\ref{conv}).

\subsection{Equations of Motion}
The equations of motion for vertex $i$ of the deformable particle are
\begin{equation}
m\frac{d^2\vec{r}_i}{dt^2} =  -{\vec \nabla}_iU + {\vec F}_f^i,
\label{eq: equation_of_motion}
\end{equation}
where $m=M/N_v$ is the mass of each vertex and $M$ is the total mass of the droplet. We integrate eqn~(\ref{eq: equation_of_motion}) using a modified velocity-Verlet numerical integration scheme with time step $\Delta t = 10^{-4}t_0$. The droplet is initialized as a regular polygon of $N_v$  sides with area $A_{\rm eq}$. We then set the edge lengths equal to their equilibrium values $l_{\rm eq} = \sqrt{4A_{\rm eq}\tan\left(\pi/N_v\right)/N_v}$. At the start of the simulation, we place the droplet at rest at a distance of $5D_0$ from the center of the obstacle to allow it to reach $v_{f\infty}$ before it collides with the obstacle.

\section{Results}
\label{results}
\begin{figure}
\centering
\includegraphics[width=0.5\linewidth]{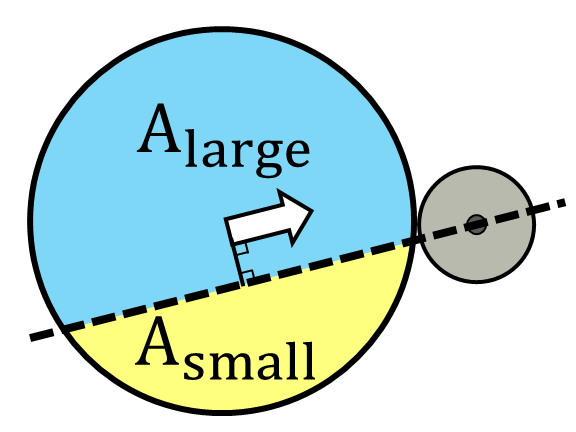}
    \caption{Illustration of the definition of the symmetry parameter $S$. When a droplet first contacts an obstacle (shaded gray), a dividing line (dashed line) is drawn through the droplet to form two regions with areas $A_\text{large}$ (shaded blue) and $A_\text{small}$ (shaded yellow). The dividing line is parallel to the center of mass velocity vector of the droplet (large white arrow) and passes through the center of the obstacle.}
    \label{fig:symmmetry}
\end{figure}

We seek to understand the key physical properties that determine droplet breakup as a droplet collides with a circular obstacle. All of the data concerning the droplet (i.e. the area, symmetry, and velocity) is collected at the frame of the first contact between the droplet and the obstacle.  An important parameter governing droplet breakup is the offset between the droplet trajectory and the obstacle. Collisions are commonly characterized using the impact parameter
\begin{equation}
    b = \left|\frac{\vec{r}_{cc} \times \vec{v}}{|\vec{v}|}\right|,
\end{equation}
\noindent where $\vec{v}$ is the velocity of the center of mass of the droplet and $\vec{r}_{cc}$ is the vector joining the center of the obstacle to the center of mass of the droplet. This definition does not capture the droplet shape when it collides with the obstacle. Thus, we define the symmetry parameter $S$ that better captures the shape of the droplet while retaining the information contained in the impact parameter.
$S$ is defined using the observed droplet area relative to a centerline at the time of the collision as shown in Fig.~\ref{fig:symmmetry}.  The centerline passes through the center of the obstacle and is parallel to each droplet's center of mass velocity when that droplet first makes contact with the obstacle.  The symmetry parameter is defined using the two subareas $A_\text{large}$ and $A_{\text{small}}$ above and below the centerline, where the total area is $A=A_\text{large}+A_\text{small}$:
\begin{equation}
    S=1-\frac{A_{\text{large}}-A_{\text{small}}}{A}.
    \label{symdef}
\end{equation}
$S=1$ indicates a perfectly symmetric collision, and $S=0$ indicates a collision where all of the droplet is on one side of the obstacle.  Droplet breakup requires $S>0$, otherwise the droplet slides around the obstacle. $S$ has an uncertainty of $\pm 0.088$ due to noise in measuring the areas.  We emphasize that $S$ is measured at the moment the droplet first contacts the obstacle and we do not focus on the subsequent evolution of $A_{\rm large}$ and $A_{\rm small}$.

In Sec.~\ref{separator_regions}, we will show that we can identify distinct regions in the parameter space where droplets break up versus where droplets do not break up using the symmetry parameter $S$ and the combined control parameter $\mathrm{Ca} \widetilde{A} \widetilde{z}$, where $\widetilde{A}=A/R^2$ is the nondimensional droplet area and $\widetilde{z}=z/R$ is the nondimensional chamber thickness. One can also nondimensionalize the geometrical parameters using the diameter of the undeformed droplet $D_0$, resulting in the nondimensional quantities: $z/D_0$ and $A/D_0^2$.  However, since the droplet area is approximately constant during the experiments (until it breaks), $A \approx \pi D_0^2/4$, which makes $A/D_0^2 \approx \pi/4$. In this case, the important nondimensional parameters become $z/D_0$ and $R/D_0$ (i.e. the obstacle size relative to the droplet size). We prefer to non-dimensionalize using $R$ instead of $D_0$ because the droplets are often deformed slightly from circular at the moment they first touch an obstacle, making direct measurements of $A$ easier in experiments, rather than $D_0$.

We find that the boundary separating the break up and no break up regimes follows a power-law relation in $S$ and $\mathrm{Ca} \widetilde{A} \widetilde{z}$. In Sec.~\ref{breakup number}, we use the power-law scaling relation to construct a dimensionless breakup number ${\rm Bk}$, where ${\rm Bk} \gg 1$ indicates that the droplet will break up and ${\rm Bk} \ll 1$ indicates that the droplet will not break up. In Sec.~\ref{daughter_sizes}, we focus on droplets that undergo breakup and determine how the ratio of the daughter droplet areas depends on $S$. Finally, in Sec.~\ref{neck_thickness}, we measure the minimum neck thickness that a droplet can sustain without breaking up and use these measurements to validate the break up model employed in the deformable particle model simulations.

\subsection{Parameter Regimes for Droplet Breakup}\label{separator_regions}

\begin{figure*}[ht]
    \centering
\includegraphics[width=0.8\linewidth]{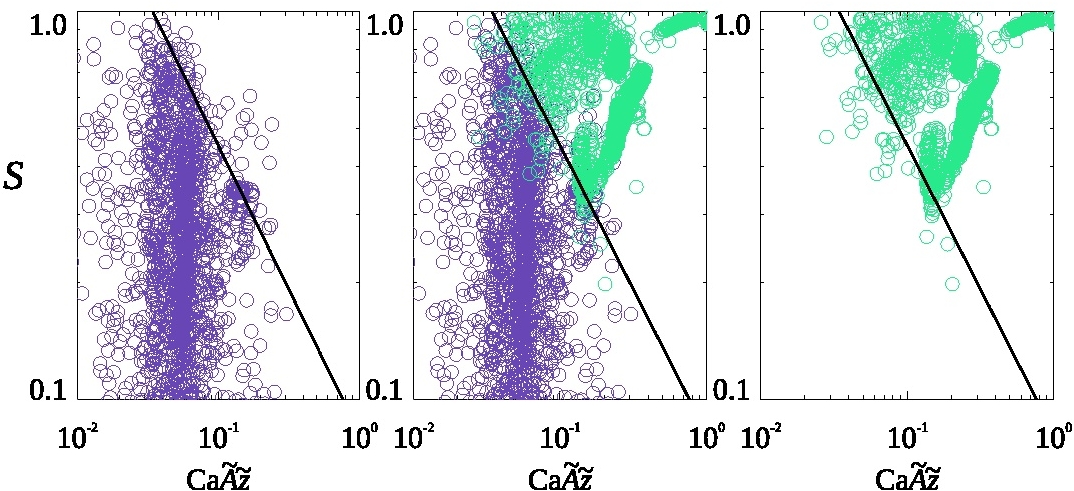}
    \caption{The symmetry parameter $S$ plotted versus ${\rm Ca} \widetilde{A} \widetilde{z}$ for all droplet collisions separated into those for which the droplets (Left) break up, (Right) do not break up, and both overlaid (Middle). The separating line (black dashed line) is given in eqn~(\ref{finalform}) with power-law scaling exponent $\beta = -0.74$.}
    \label{fig:Main}
\end{figure*}

In Fig.~\ref{fig:Main}, we separate the experimental data for droplet-obstacle collisions into two sets in the parameter space of $S$ versus ${\rm Ca} \widetilde{A} \widetilde{z}$: (Left) droplets that break up and (Right) droplets that do not break up. These results show that droplet break up involves a tradeoff between $S$ and ${\rm Ca} \widetilde{A} \widetilde{z}$; for example, a droplet that collides with the obstacle off-center ($S < 1$) can be made to break up by increasing the velocity, since Ca $\sim v$.  The two clouds of experimental data are best separated by a power-law scaling form:
\begin{equation}
    S_c=\alpha \left(\mathrm{Ca}\widetilde{A}\widetilde{z}\right)^{\beta}, \label{finalform}
\end{equation}
where the prefactor $\alpha \approx 0.083$ and the power-law scaling exponent $\beta \approx -0.74$. Since $S_c$ scales with ${\rm Ca}$, droplets are more likely to break up when they have a large velocity, are immersed in a fluid with large viscosity, and have small surface tension. 

Eqn~(\ref{finalform}) shows that larger droplets are also easier to break, i.e. a larger object feels more pressure when pushed against a smaller object; the smaller object is effectively sharper for larger $\widetilde{A}$.  Our data have $\widetilde{A} \sim O(1)$.  When $\widetilde{A} \ll1$, we expect different behavior than that shown in Fig.~\ref{fig:Main} because the obstacle will appear more like a wall, which is a situation that has been previously studied.\cite{fu_dynamics_2011}  

Additionally, droplets in thicker chambers are more likely to break up. Between the parallel plates of the sample chamber, the droplet interface curves between the top and bottom surface with a radius of curvature $\sim z/2$.\cite{desmond13}  This small curvature increases the Laplace pressure inside the droplet, $\Delta P \sim \gamma/z$.  Thus, thicker droplets are ``softer'' and more easily deform and break.

\begin{figure}
    \centering
    \includegraphics[width=0.95\linewidth]{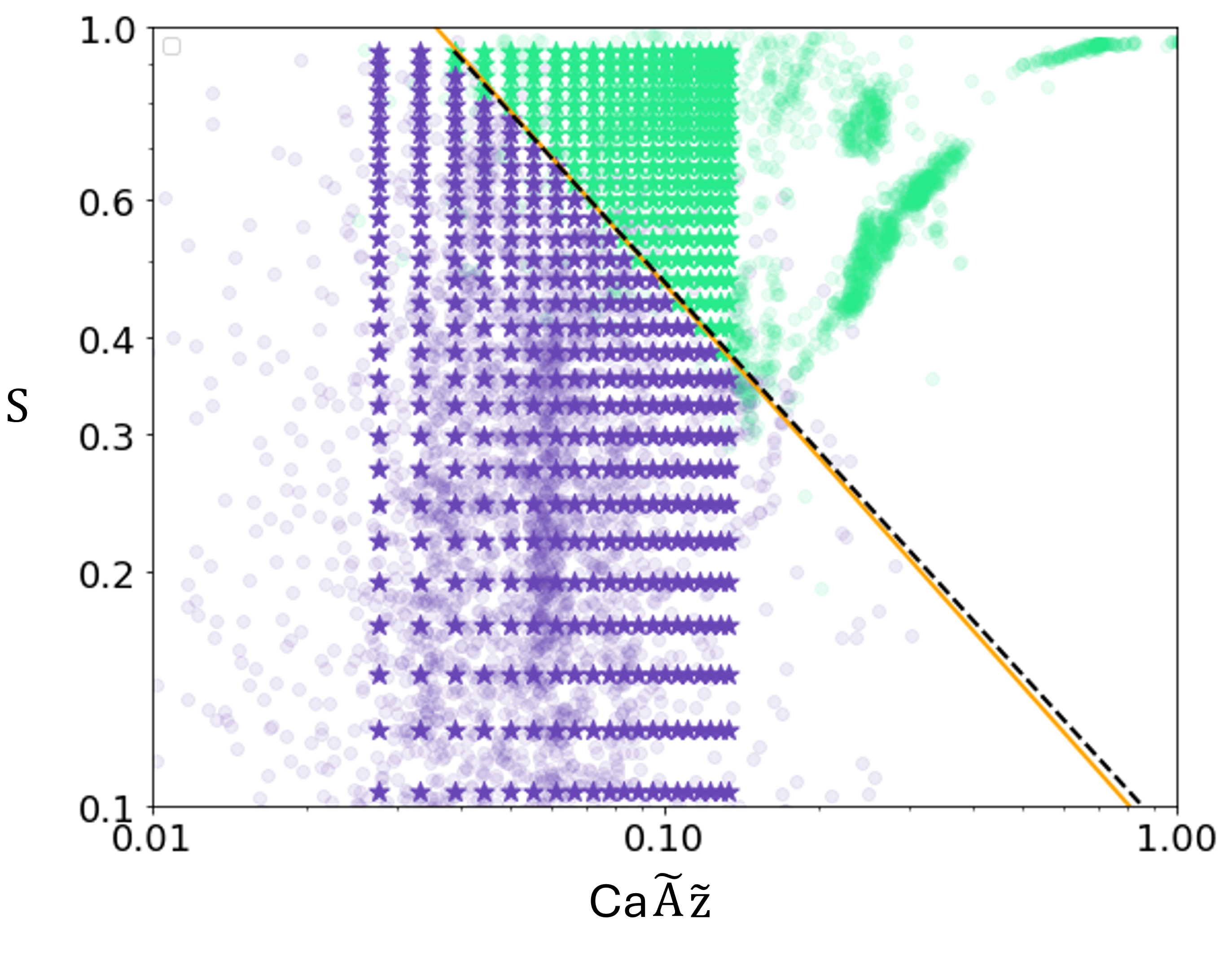}
    \caption{Results from the deformable particle model (DPM) simulations (stars) showing droplets that break up (upper right; green) and do not break up (lower left; violet) as a function of $S$ and ${\rm Ca} \widetilde{A}\widetilde{z}$ overlaid on the experimental data from Fig.~\ref{fig:Main}. We set $\widetilde{z}=1$ for the simulation data. The best fit lines that separate the droplets that break up and do not break up have slopes $-0.72$ (black dashed; simulations) and $-0.74$ (orange solid; experiments).}
    \label{fig:Mainsim}
\end{figure}

In the experiments, the surface tension is fixed and the viscosity is varied between two values, 50 cSt and 100 cSt. Consequently, in experiments we varied the capillary number primarily by changing the droplet velocity. In contrast, the simulations allow independent changes of the surface tension and viscosity. By varying these parameters independently, we show in Appendix~C that $S_c \sim \left(\widetilde{\mu}/\widetilde{\gamma}\right)^\beta$ with $\beta \approx -0.72$, which is in agreement with the experimental results as shown in Fig.~\ref{fig:Mainsim}. Note that because the simulations are two-dimensional, the chamber thickness $\widetilde{z}$ cannot be varied. Hence, we set $\widetilde{z}=1$ for the simulation data in Fig.~\ref{fig:Mainsim}. We performed a sweep over a range of values of $S$ and $\mathrm{Ca}\widetilde{A}\widetilde{z}$, noting whether the droplet breaks up or not for each $(S, \mathrm{Ca}\widetilde{A}\widetilde{z})$ pair. We then hold $\mathrm{Ca}\widetilde{A}\widetilde{z}$ fixed and increase $S$ until we find the largest value $S_c$ for which the droplet does not break up. Repeating this for all of the $\mathrm{Ca}\widetilde{A}\widetilde{z}$ values in our simulation gives us a set of points that form the separation boundary between regions of no-breakup and break up. We find that this separation curve is best described by the power law
\begin{equation}
    S_c = 0.089\left(\mathrm{Ca}\widetilde{A}\widetilde{z}\right)^{-0.72},
\end{equation}
which is excellent agreement with the separation curve obtained from experiment, as can be seen in Fig.~\ref{fig:Mainsim} and by direct comparison to eqn (\ref{finalform}).

\subsection{Breakup Number} \label{breakup number}

To determine the power-law exponent in eqn~(\ref{finalform}), we use the maximum likelihood method.\cite{nelson}  We start by assuming an equation for the dividing line of the form:
\begin{equation}
    S_c=\alpha \left(\mathrm{Ca}A^{\epsilon_1}R^{\epsilon_2}z^{\epsilon_3}\right)^\beta,
    \label{leadup}
\end{equation}
where $\beta$ is the power-law exponent.  Given that $A, R$, and $z$ all have dimensions related to length, we have the constraint $2\epsilon_1+\epsilon_2+\epsilon_3=0$ such that eqn (\ref{leadup}) is nondimensional.  We next define the ``breakup number" 
\begin{equation}
\text{Bk}=\alpha^{1/\beta}\mathrm{Ca}A^{\epsilon_1}R^{\epsilon_2}z^{\epsilon_3}/S^{1/\beta},
\label{bkeqn}
\end{equation}
to quantify the distance from the dividing line $S_c$. With these definitions, droplets that break up will have large values of Bk, and droplets that do not break up will have ${\rm Bk} \rightarrow 0$.

We apply the maximum likelihood method for separating the experimental data for which droplets break up 
and droplets do not break up because the experimental data is noisy near $S_c$.  As shown in Fig.~\ref{fig:Main}, some droplet-obstacle collisions are observed on the ``wrong'' side of the dividing line.  There are examples in the data set close to $S_c$ where for similar experimental parameters, some droplets break and others do not.  One reason for this behavior could be small uncertainties in the experimental measurements of velocity or $S$ stemming from the image analysis. However, these quantities are defined at the moment a droplet first contacts the obstacle, and a more likely concern is that droplets can and do change their speed and direction of motion as they interact with the obstacle.  In addition, other droplets near the given droplet-obstacle collision may influence the flow of oil around the droplet, again changing the behavior of the droplet as it interacts with the obstacle. Therefore, for the experimental data, we consider breakup as a probabilistic process, and Bk$=1$ corresponds to the case where droplets are equally likely to break or not.

To mathematically implement the breakup probability, we define the breakup characteristic $k$ as:
\begin{equation}
k=\begin{cases} 
      1 & \text{breakup} \\
      0 & \text{no breakup}.
   \end{cases}
    \label{likelihood1}
\end{equation}
We then define the probability of observing outcome $k$, using $x = (\text{Bk}_{\alpha,\beta})^{1/w}$ (where the subscripts indicate that Bk is a function of $\alpha$ and $\beta$) and the function
\begin{equation}
P(x,k)=\begin{cases} 
      x/\left(1+x\right) & k=1 \\
      1/\left(1+x\right) & k=0,
   \end{cases}
    \label{likelihood2}
\end{equation}
which means that for $\text{Bk}_{\alpha,\beta} \gg 1$ break up is likely $[P(k=1) \rightarrow 1]$ and for $\text{Bk}_{\alpha,\beta} \ll 1$ no break up is likely $[P(k=0) \rightarrow 1]$. $P(x,k)$ is a sigmoid function of $\ln(\textrm{Bk})$, where $w$ is the width of the sigmoid.  We then define the likelihood $L$ of observing the data by a product over all droplet-obstacle collision events $i$ as:
\begin{equation}
\label{likelihood3}
    L_{\alpha,\beta,w}=\prod_i P(({\text{Bk}^i_{\alpha,\beta})^{1/w},k_i}).
\end{equation}
Terms in the product are close to $0$ for data points with a low probability of the actual outcome, and close to $1$ when the prediction matches the actual outcome. Therefore, incorrectly chosen $\alpha$ and $\beta$ dramatically reduce the likelihood due to the contributions from many incorrectly assigned points. In contrast, optimal $\alpha$ and $\beta$ will create many more matches and a much larger total likelihood.  $w$ accounts for the width of the region where droplets have an intermediate chance of break up.  Maximizing the logarithm of the likelihood~\cite{nelson} yields power-law exponent $\beta = -0.74$ and $\alpha \approx 0.083$.

We also use the maximum log-likelihood method to calculate the exponents $\epsilon_{1}$, $\epsilon_2$, and $\epsilon_3$ in eqn~(\ref{bkeqn}) that are subject to the constraint $2 \epsilon_1 + \epsilon_2 + \epsilon_3=0$.  We find that $S_c \sim A^1 z^1 R^{-3}$ as shown in Fig.~\ref{fig:Main}. As an additional check, we allowed the power-law scaling exponent $\epsilon_\mu$ for $\mu$ in eqn~(\ref{bkeqn}) to vary. Maximizing the log-likelihood returned $\epsilon_\mu=1$, confirming the Ca dependence for $S_c$.
We determine the uncertainty of $\alpha$, $\beta$, and $w$ using the bootstrapping method (with $100$ samples of 2,528 randomly selected data points). We find $\alpha=0.083\pm 0.006$, $\beta=-0.74 \pm 0.03$, and $w=0.098\pm0.006$.

We now present an argument to interpret the observed value of the power-law exponent $\beta \approx -0.74$ in eqn~(\ref{finalform}).  We will show that an exponent of $\beta=-3/4$ corresponds to a conserved geometric quantity during the collision, the minimum height of $A_{\rm small}$. Consider droplet-obstacle collisions with fixed velocity, geometry, and fluid properties, but varying $A$.  In this case, droplet-obstacle collisions on the dividing line in Fig.~\ref{fig:Main} are described by $S_c = (A/\Lambda)^\beta$, where all of the other parameters are subsumed into $\Lambda$ with units of area.  We can then use eqn~(\ref{symdef}) to rewrite $S_c$ in terms of the two droplet subareas:
\begin{equation}
    S_c=1-\frac{A_\text{large}-A_\text{small}}{A} = \left( \frac{A}{\Lambda} \right)^{\beta}.
    \label{asmall}
\end{equation}
Solving for $A_\text{small}$ gives:
\begin{equation}
    A_\text{small} = \frac{1}{2}A^{1+\beta} \Lambda^{-\beta}.
    \label{symmatheq}
\end{equation}

\begin{figure}
    \centering
\includegraphics[width=0.75\linewidth]{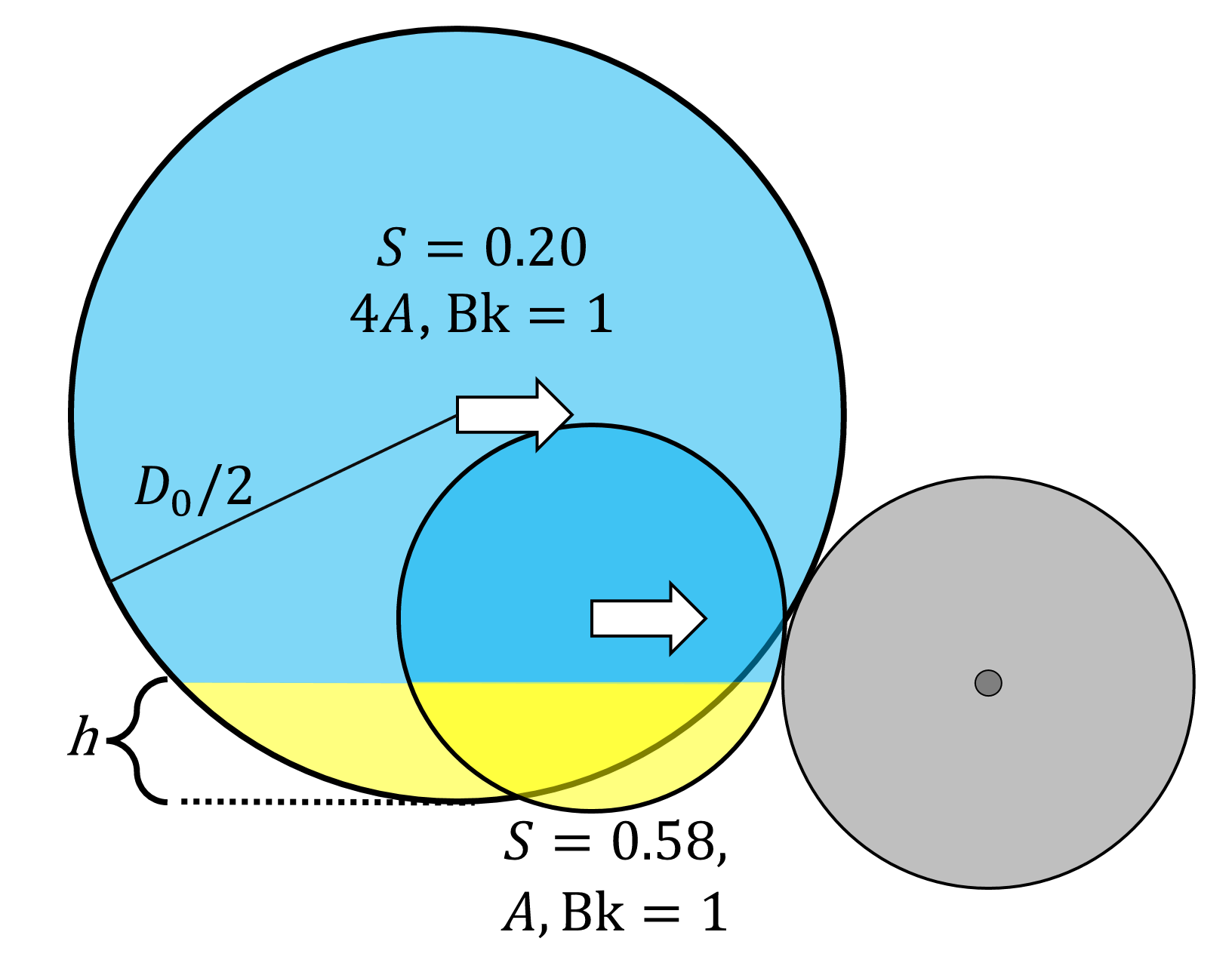}
    \caption{Image of two droplets colliding with an obstacle (shaded gray). The droplets have equal velocities ($v=1.0$~mm/s), identical chamber geometries, the same fluid parameters, but one droplet is twice the radius of the other. The smaller droplet has $A=0.02$~mm$^2$. The two droplets have $\text{Bk}=1$, and the smaller area below the symmetry line (in yellow) of the larger droplet is $1.4$ times that of the smaller droplet. The droplet radius $D_0/2$ and height of the smaller circular section $h$ are indicated.}
    \label{fig:symmathd0}
\end{figure}

A comparison of droplet-obstacle collisions for two droplet areas is shown in Fig.~\ref{fig:symmathd0}, where the yellow-shaded circular segments have area $A_{\text{small}}$.  The figure defines the length scale $h$ and droplet diameter $D_0$, and in the limit $h \ll D_0$ one can show $A_{\text{small}} \sim D_0^{1/2} h^{3/2}$. We can relate the diameter and area of the droplet using $D_0 \sim A^{1/2}$ and express eqn~(\ref{symmatheq}) for $A_{\rm small}$ in terms of $h$:
\begin{equation}
    h \sim A_{\text{small}}^{2/3} A^{-1/6}.
    \label{hsima}
\end{equation}
Next, consider the case $h \sim A^0$, where $h$ has a fixed value that is independent of $A$ for a droplet-obstacle collision at $S_c$.  With these assumptions, we substitute eqn~(\ref{symmatheq}) into eqn~(\ref{hsima}) to obtain
\begin{equation} 
    A^0 \sim (A^{1 + \beta})^{2/3} A^{-1/6}. 
\end{equation}
To ensure that $h$ is independent of $A$, we must set $\beta = -3/4$.  The power-law exponents $\beta = -0.74$ obtained in the experiments and $\beta = -0.72$ obtained in the simulations are in excellent agreement with this scaling analysis. Using eqns (\ref{symmatheq}) and~(\ref{hsima}), we can also show how the small circular segment height $h$ scales with the obstacle radius $R$, sample thickness $z$, and Ca: 
\begin{equation}
    h \sim \sqrt{\Lambda} \sim \frac{R}{\sqrt{\widetilde{z} \mathrm{Ca}}}.
    \label{hequation}
\end{equation}
Thus, our results in Figs.~\ref{fig:Main} and~\ref{fig:Mainsim} suggest that the separating curve $S_c$ depends on the droplet area $A$ such that $h$ is constant, which yields the scaling behavior in eqn~(\ref{hequation}).

\begin{figure}
    \centering
\includegraphics[width=0.99\linewidth]{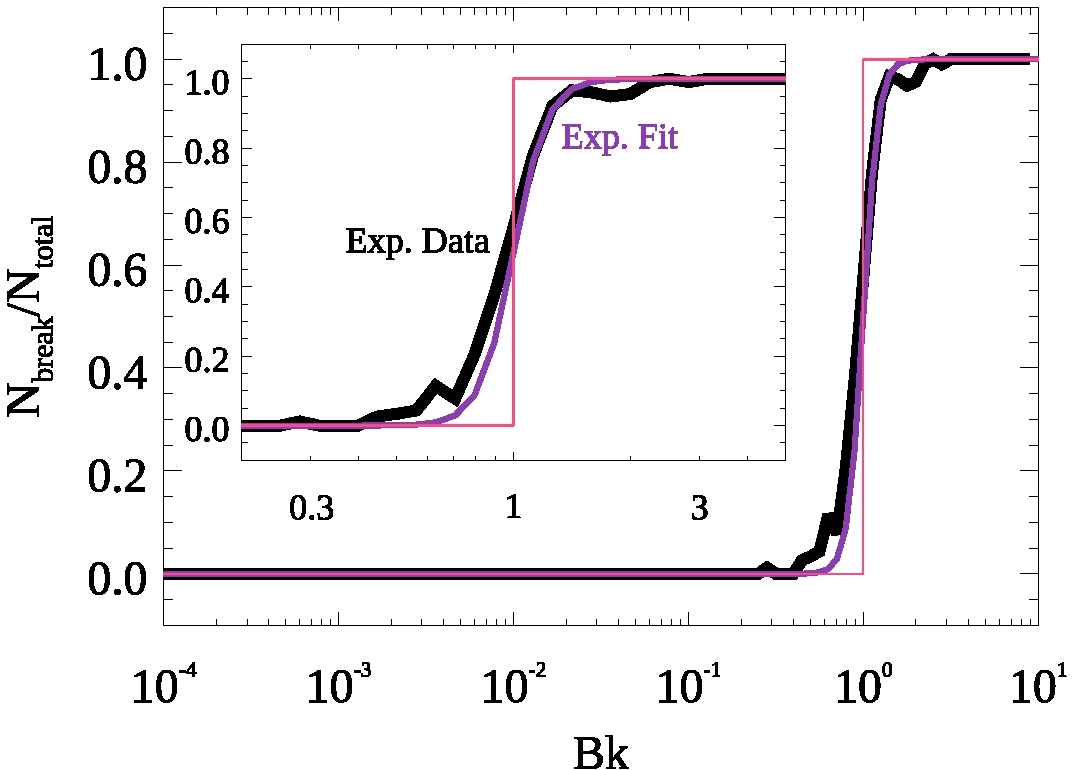}
    \caption{The fraction $N_{\rm break}/N_{\rm total}$ of droplet-obstacle collisions in the experiments (black solid line) that yield droplet break up plotted versus $\text{Bk}$. eqn~(\ref{likelihood2}) (violet solid line) and the step function $\Theta({\rm Bk})$ (pink solid line) are also shown. (inset) A close-up of $N_{\rm break}/N_{\rm total}$ in the main panel near ${\rm Bk} =1$.}
    \label{fig:probability}
\end{figure}

We now consider experimental measurements of Bk. Using eqn~(\ref{bkeqn}), we can write the breakup number as:
\begin{equation}
\text{Bk}=28\mathrm{Ca}\widetilde{A}\widetilde{z}S^{-4/3}, 
\label{Bkequation}
\end{equation}
where $\text{Bk}<1$ indicates that the droplets are not likely to break up, and $\text{Bk}>1$ indicates that the  droplets are likely to break up. In Fig.~\ref{fig:probability}, we plot the fraction $N_{\rm break}/N_{\rm total}$ of droplet-obstacle collisions that yield droplet break up as a function of ${\rm Bk}$. The probability is a sigmoid function with a narrow width: the range of Bk over which the probability grows from $5$\% to $95$\% is only a factor of $\approx 2$ in ${\rm Bk}$. The results in Fig.~\ref{fig:probability}, which extend over six orders of magnitude in Bk, illustrate that we have identified the key parameters that determine droplet break up during single droplet-obstacle collisions.

\subsection{Area Ratio of Daughter Droplets} \label{daughter_sizes}

\begin{figure}
    \centering
\includegraphics[width=0.95\linewidth]{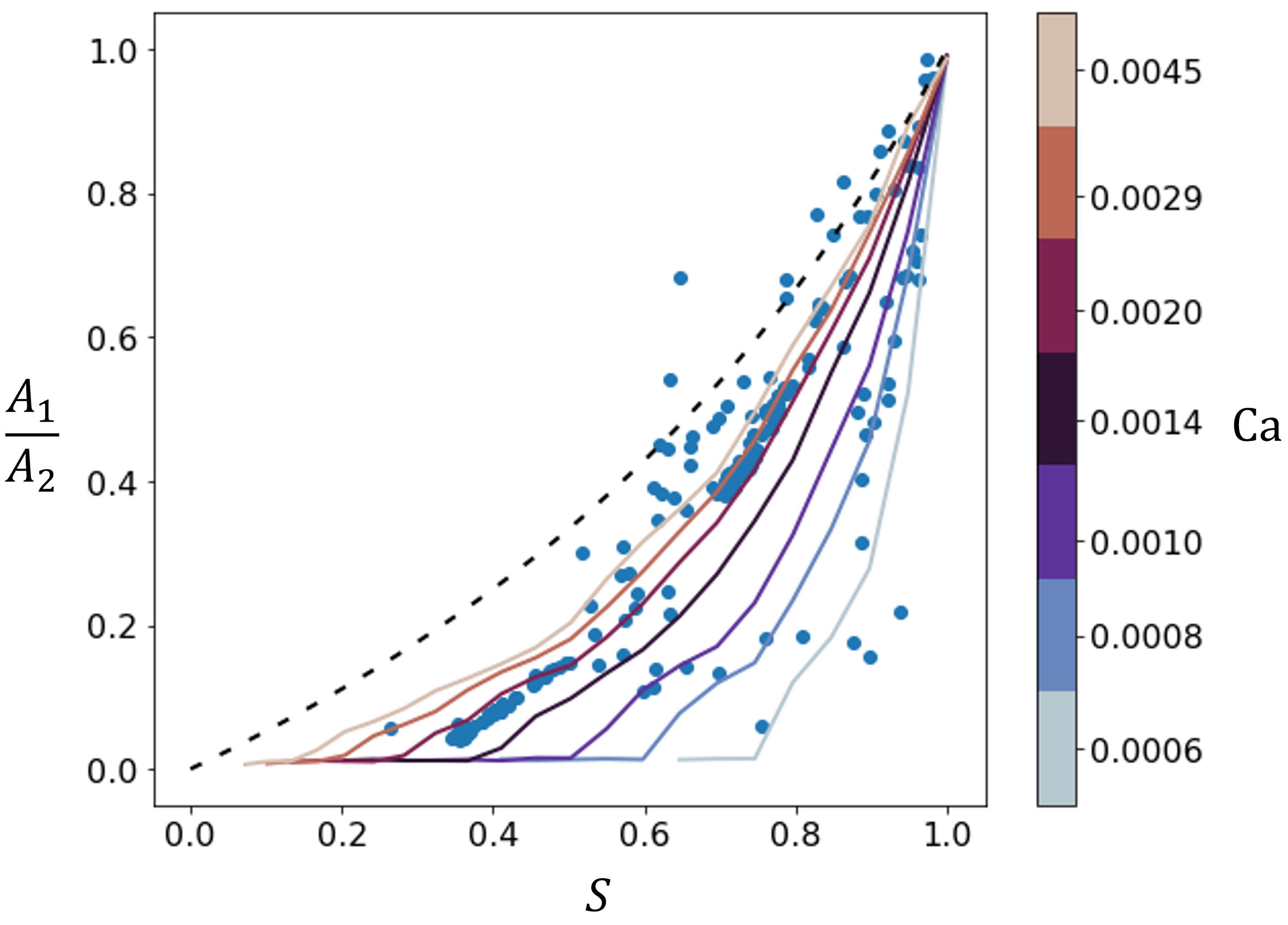}
    \caption{The daughter droplet area ratio $A_1/A_2$ for droplet-obstacle collisions that yield droplet break up plotted verses the symmetry $S$ of the collision in experiments (blue filled circles).  We also show eqn~(\ref{DDR}) (dotted black line), where the droplet breaks along the collision axis with no fluid exchange, $A_1 = A_{\rm small}$, and $A_2 = A_{\rm large}$. The simulation data (solid lines) are shaded according to the value of Ca from ${\rm Ca}=0.0006$ to $0.0045$. Two linear clusters of data are observed; these correspond to experiments where droplets collide reproducibly over a narrow range of $S$, leading to $A_1/A_2  \sim S$.}
    \label{fig:DDR}
\end{figure}

As noted above, all of the experiments are in the low Reynolds number regime, Re $\leq O(10^{-2})$.  In this regime, we find that nearly all droplet break up events  result in the formation of only two daughter droplets. If we denote the area of the smaller daughter droplet as $A_1$ and that of the larger daughter droplet as $A_2$, we can define the daughter droplet area ratio $0 < A_1/A_2 <1$, which depends on Ca and $\widetilde{A}$ of the original droplet. We contrast these areas with $A_{\rm{small}}$ and $A_{\rm{large}}$, which are defined for each droplet at the first contact between the droplet and the obstacle. In off-centered collisions ($S<1$), we find that $A_1 < A_{\rm{small}}$ and $A_2 > A_{\rm{large}}$ (with $A_1+A_2 = A_{\rm{small}} + A_{\rm{large}}$ from mass conservation), because the larger lobe partially slides along the obstacle prior to breakup, dragging the smaller lobe with it, and thereby redistributing area before break up. However, when ${\rm Ca} \gg 1$, we expect that the timescale for mass redistribution is much larger than the timescale for droplet deformation (and subsequent break up). In this limit, $A_1 \to A_{\rm{small}}$ and $A_2 \to A_{\rm{large}}$, and thus
\begin{equation}
    \lim_{\mathrm{Ca}\gg 1}\frac{A_1}{A_2} =\frac{A_{\rm{small}}}{A_{\rm{large}}} = \frac{S}{2-S}.
    \label{DDR}
\end{equation}
In Fig.~\ref{fig:DDR}, we plot the daughter droplet area ratio $A_1/A_2$ versus $S$ for droplet-obstacle collisions that yield droplet break up from experiments and simulations. We observe that all of the simulation data and nearly all of the experimental data occur below this limiting curve. However, given the presence of continuous phase fluid flow driven by other nearby droplets, it is possible for some of the droplet fluid to move into the smaller lobe such that $A_1/A_2$ exceeds the limiting prediction. Other confounding influences on $A_1/A_2$ in the experiment are droplets that change velocity while they are in contact with the obstacle, and droplets that are pre-deformed upon the initial contact with the obstacle (due to the motion of other nearby droplets). 

We also show $A_1/A_2$ versus $S$ in Fig.~\ref{fig:DDR} from the simulations for a range of Ca. For the simulations, we find that $A_1/A_2$ approaches $S/(2-S)$ in the large-Ca limit in agreement with eqn~(\ref{DDR}). In particular, $A_1/A_2=1$ when $S=1$. In this case, the two lobes are identical and there is no fluid redistribution across the lobes during break up for all Ca. For $A_1/A_2$ to reach $S/(2-S)$ for $S<1$, Ca must progressively increase as $S$ decreases. 

\subsection{Droplet Neck Thickness} 
\label{neck_thickness}

\begin{figure}
    \centering
\includegraphics[width=0.99\linewidth]{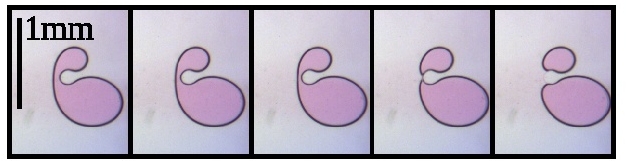}
    \caption{Experimental snapshots of a droplet breaking up. The images are taken at intervals of 1/60~s to show the neck thinning and breaking.  The image analysis places the exact moment of breakup between the 3rd and 4th frames from the left.}
    \label{fig:neckpics}
\end{figure}

We also examine the minimum neck thickness $d_{\text{neck}}$ (as defined in Sec.~\ref{mesoscale}) attained during droplet–obstacle interactions, considering both droplets that undergo breakup and those that do not. We display the evolution of the neck of a breaking droplet from experiments in Fig. \ref{fig:neckpics}.

The neck is experimentally measured in the single video frame immediately before the frame where the droplet has broken into two daughter droplets (for example, the middle image of Fig.~\ref{fig:neckpics}).  Image analysis identifies the perimeter pixels of the droplet. We assign each pixel an index $i$ ordered sequentially along the perimeter.  We then compute the distance between each pixel pair $(i,j)$.  In the space $(i,j)$ the line $i=j$ forms the global minimum as the distance from each pixel to itself is zero.  There is also a local minimum away from the diagonal that corresponds to the minimum distance between two pixels with indices $i$ and $j$ far apart (along the contour of the perimeter), which identifies the neck thickness and the two specific pixels $i$ and $j$ defining the neck.

\begin{figure}
    \centering
\includegraphics[width=0.99\linewidth]{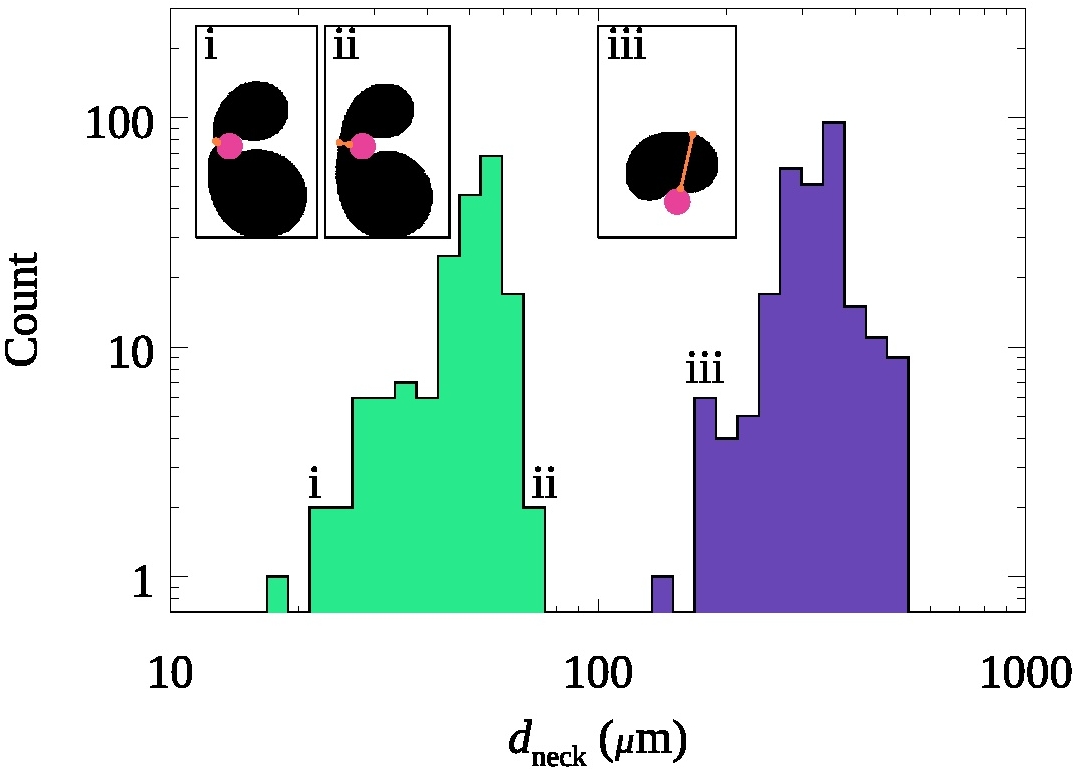}
    \caption{Frequency distribution for the minimum neck thickness $d_{\rm neck}$ of droplets that (green) do and (violet) do not break up in experiments. (insets) Example droplets (i-iii) for regions (i-iii) of $d_{\rm neck}$ marked on the histograms, with fluid flow from right to left, obstacle radius $R=85$~$\mu$m (shaded pink), and neck outline (orange solid line). The data shown are for the standard experimental parameters given in Sec.~\ref{sec:exptmethods}.}
    \label{fig:neck}
\end{figure}

In Fig.~\ref{fig:neck}, we show the frequency distribution of the thinnest neck $d_{\rm neck}$ observed in the experiments. We find that droplets which undergo breakup never appear in their final recorded frame with neck thicknesses larger than $d_{\rm neck} \approx 70~\mu$m (green histogram), while droplets that do not break always have $d_{\rm neck} \gtrsim 140\,\mu\mathrm{m}$ (violet histogram). Based on studies of instability-mediated fluid filament breakup, we expect that there is a critical neck thickness $d_{\rm min-neck}$ below which the droplet will always break up \cite{shi_cascade_1994}. The histogram of neck thicknesses for non-breaking droplets sets a ceiling on $d_{\rm min-neck}$: droplets with $d_{\rm neck} < 140$~$\mu$m always break up.  It is possible if we observed a much larger number of droplets, we would find rarer cases of non-breaking droplets with slightly thinner necks that would reduce the ceiling. The value of $d_{\rm neck-thin}$ is expected to depend on factors such as the droplet size, chamber thickness, and possibly the surface tension; we have insufficient data to determine this dependence so we simply estimate $d_{\rm min-neck} \approx 140$~$\mu$m.

The green histogram (corresponding to droplets that break) depends on the frame rate of the camera (60 frames per second).  Droplet neck thinning likely has a rapid stage at the end \cite{shi_cascade_1994} but nonetheless is expected to be a continuous process, so a higher speed camera will reduce the observed $d_{\rm neck}$ as seen one video frame before breakup.  The green histogram can thus be interpreted as an observation that some droplets with a neck thickness up to $\sim 70$~$\mu$m can thin to zero on a time scale of 1/60~s or faster.  The location and breadth of this histogram for breaking droplets reflects the finite frame rate of the camera, which captures droplets at different stages of neck-thinning prior to breakup. As a result, we expect that increasing the frame rate would decrease the width and location of the green histogram.

In the simulations, we set $d_{\rm min\text{-}neck} = 0.167 D_0$. Using the experimentally observed median diameter of droplets that undergo breakup, $D_0 = 481~\mu$m, this corresponds to $d_{\rm min\text{-}neck} \approx 80.3~\mu$m, which is consistent with the experimental bounds.

\section{Conclusions}
\label{conclusions}

We carried out coordinated experiments and simulations of quasi-two-dimensional deformable droplets flowing in microfluidic chambers that collide with cylindrical pillars. For some conditions, droplet-obstacle collisions give rise to droplet break up. For others, droplets collide with the obstacle, but slide around it, and do not break up. Break up is influenced by the interplay between the viscous surface stresses that deform the droplet and the surface tension that resists deformation, which is quantified by the capillary number Ca in eqn~(\ref{eqn:capillary}).  We find that droplet break up also depends on several geometrical parameters.  For example, larger droplets relative to the obstacle are more likely to break up.  Droplets in thinner sample chambers are less likely to break up, as a result of the larger internal Laplace pressure that resists droplet deformation required for break up.  In addition, we show that the symmetry of the droplet trajectory relative to the center of the obstacle influences droplet break up, i.e. head-on collisions with $S=1$ maximize the likelihood of break up. The results for droplet break up can be collapsed using the nondimensional breakup number Bk, which is proportional to $\mathrm{Ca}\widetilde{A}\widetilde{z} S^{4/3}$, and $\widetilde{A}$ and $\widetilde{z}$ are the droplet area and sample chamber thickness normalized by the obstacle radius $R$.  Bk accurately predicts the likelihood of droplet breakup in experiments over six orders of magnitude of Bk, with a narrow region of Bk near $\text{Bk}=1$ where the droplet break up probability transitions from zero to one with increasing Bk. 

The experimental results are also verified through discrete element method simulations using the deformable particle model with line tension in 2D. We demonstrate that incorporating a geometric criterion for droplet break up related to the neck thickness is sufficient to capture the droplet break up statistics. In particular, we find that the characteristic symmetry for which a droplet undergoes breakup is $S_c = \alpha\left(\mathrm{Ca}\widetilde{A}\widetilde{z}\right)^\beta$, which is the same functional form observed in experiments, and the power-law exponent $\beta\approx -0.72$ is in close agreement with experiments. Further, the geometric break up criterion implemented in the simulations is consistent with the experimental observations that droplets whose neck thickness is above a threshold value do not break up.

Our work gives additional insight into the daughter droplets formed after break up. We observe that single obstacle-induced droplet breakup events predominantly produce two daughter droplets, and the resulting daughter droplet area ratio is constrained by the initial symmetry of the collision. For off-centered collisions, interfacial sliding and lobe interactions prior to break up redistribute area, causing the daughter droplet areas to deviate from those defined at first contact and the daughter droplet area ratio to be below the upper bound $A_1/A_2 = S/(2-S)$. In the limits of large capillary number and large droplet area, area redistribution is negligible and the daughter droplet area ratio approaches the upper bound, which is verified by the simulations.

These results suggest several promising future research directions.  First, we can extend our studies of droplet break up to three dimensions, where thin droplet necks are completely unstable due to surface tension. Second, droplets can coalesce, as well as break up.\cite{bremond_decompressing_2008,lai_separation-driven_2009} In future studies, we will investigate droplet coalescence in a microfluidic porous medium. With a fundamental understanding of both droplet break up and coalescence, we will be able to predict the resulting droplet size distributions as they move through the medium.\cite{Yiotis2013}  As droplets flow though a microfluidic porous medium composed of obstacles, we expect that a steady-state size distribution will be reached where break up events are balanced by coalescence events.\cite{amstad_microfluidic_2014} We can also consider the problem of fluid wetting; in real porous media flows, droplets are not perfectly dewetted to the media.\cite{woche_soil_2017} This effect may have significant effects on fluid-flow induced droplet break up and coalescence.

\section*{Author contributions}
Conceptualization: All authors
\\
Methodology: David J. Meer (Experimental). Shivnag Sista (Computational).
\\
Investigation: David J. Meer (Experimental). Shivnag Sista (Computational).
\\
Formal analysis: David J. Meer, Eric R. Weeks (Experimental). Shivnag Sista, Mark D. Shattuck, Corey S. O'Hern (Computational).
\\
Data curation: David J. Meer (Experimental). Shivnag Sista (Computational).
\\
Writing - original draft: David J. Meer (Experimental). Shivnag Sista (Computational).
\\
Writing - review \& editing: All authors
\\
Visualization: All authors
\\
Supervision: Mark D. Shattuck, Corey S. O'Hern, Eric R. Weeks
\\
Funding acquisition: Mark D. Shattuck, Corey S. O'Hern, Eric R. Weeks

\section*{Conflicts of interest}
There are no conflicts to declare.

\section*{Data availability}
Data for this article, including experimental microscopy videos and corresponding image analyses, computational data and code to recreate the figures are available in the UNC dataverse at \hyperlink{https://doi.org/10.15139/S3/CZM4AN}{https://doi.org/10.15139/S3/CZM4AN}.

\section*{Appendix A: Microfluidic Chamber Design} 
\begin{figure}[ht]
    \centering
\includegraphics[width=0.99\linewidth]{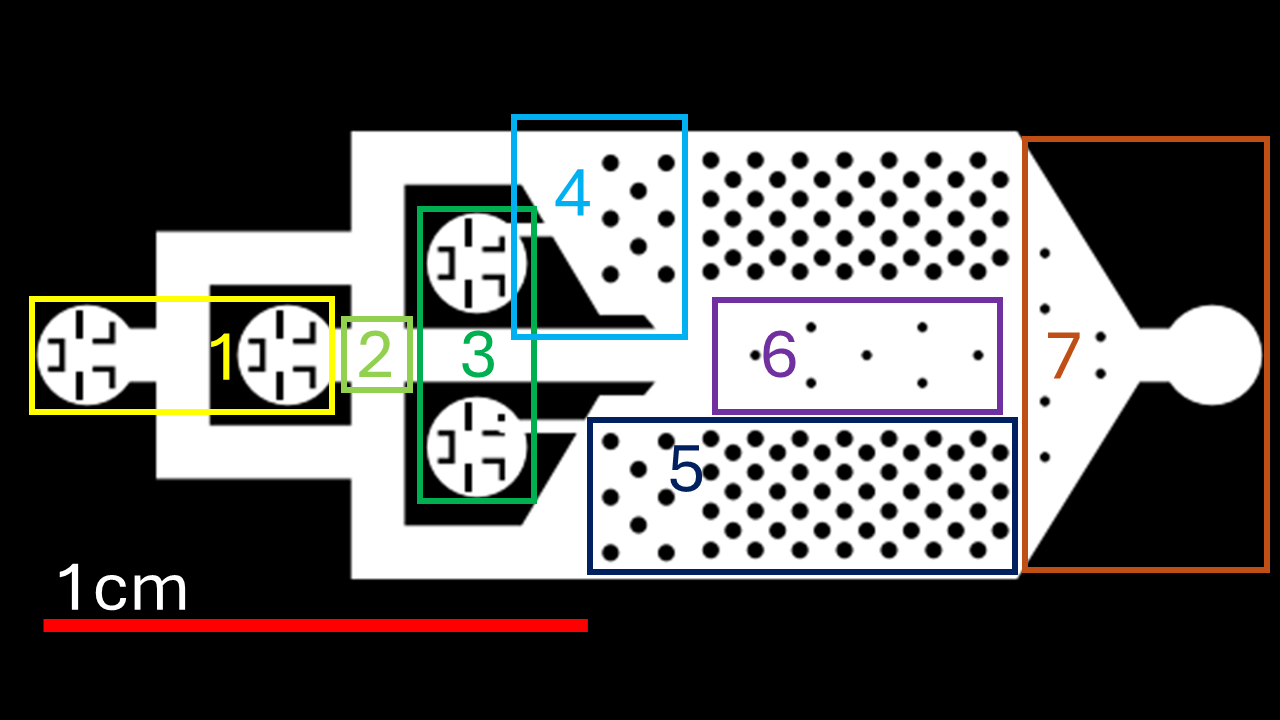}
    \caption{We label regions of the microfluidics chamber pattern. The design was converted into a photomask after it was designed in openSCAD.}
    \label{fig:appendixA}
\end{figure}
In this appendix, we describe individual sections of the microfluidics chamber pattern labeled $1$-$7$ in   Fig.~\ref{fig:appendixA}:
\begin{enumerate}
    \item  (Yellow) Inputs for the continuous oil-based phase (left) and the droplet water-based phase (right). The design within the input regions catches dust and debris brought in with the fluid.
    \item (Light green) The central region where droplet formation occurs via pinch-off of the water phase.\cite{Utada2007} The droplets then continue down the central channel, and the oil travels through all three channels.
    \item (Green) Extra oil input regions to modify the symmetry of the droplet-obstacle collisions. These inputs were not used for any of the experiments in this article, and once they were filled with oil, they did not affect the flow of the droplets.
    \item (Light blue) The oil phase expands in cross-sectional area, and slows down. This main expansion occurs before the droplet phase enters the main chamber, which ensures a minimal velocity gradient between the main observing region and central droplet channel.
    \item (Dark blue) Structural obstacles to ensure minimal wall effects, a consistent cross-sectional area of the flow, and structural stability to prevent the chamber from collapsing.
    \item (violet) The main imaging area, where obstacles have radius $R$.
    \item (Orange) The exit region, where the wall separation decreases and flow velocity increases. Therefore, no droplet behavior is recorded in this region, though the obstacles still have radius $R$.
\end{enumerate}

\section*{Appendix B: Comparing Droplet Shape and Motion from Simulations and Experiments} \label{Appendix-B}

\begin{figure}[ht]
    \centering
\includegraphics[width=0.99\linewidth]{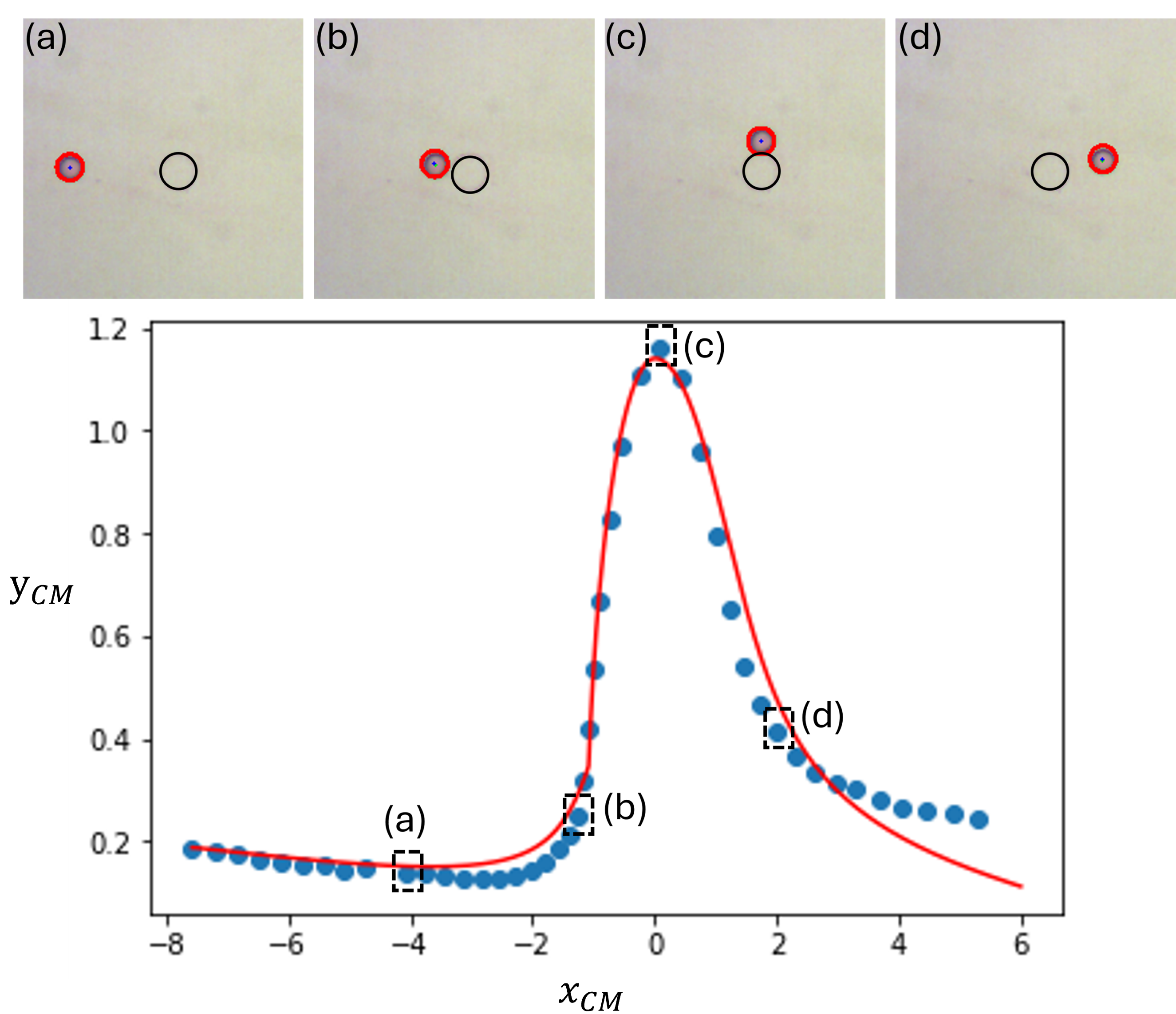}
    \caption{(a)–(d) Snapshots of a small droplet $(D_0/R = 0.78)$ flowing past an obstacle ($R=85\mu$m) in experiments with snapshots from the deformable particle model simulations overlaid in red. The droplet moves from panels (a) to (d) in $\sim 0.43$ seconds. The bottom panel shows the $y$-component of the droplet center of mass $\text{y}_{CM}$ plotted versus the $x$-component $\text{x}_{CM}$ for the simulations (solid line) and experiments (filled circles).}
    \label{fig:Traj_small}
\end{figure}

In this appendix, we show that the deformable particle model simulations (described in Sec.~\ref{sec: Fluid flow}) accurately recapitulate the flow trajectories of the droplets and the droplet shapes as they interact with the obstacles, especially in the limit of small droplets relative to the obstacles $D_0/R \ll 1$.  
In Fig.~\ref{fig:Traj_small} (a)–(d), we provide experimental images of a droplet with $D_0/R =0.78$ that does not break up with the simulation results overlaid on the experimental results. The bottom panel shows the trajectory of the droplet around the obstacle for both the experiments and simulations. In particular, we plot the $y$-coordinate of the droplet center of mass $y_{\rm CM}$ versus the $x$-coordinate $x_{\rm CM}$, where the droplet flow is in the $x$-direction. Note that for the simulations, the center-of-mass coordinates are obtained from the $x$- and $y$-coordinates of the vertices $x_i$ and $y_i$:
\begin{align}
\text{x}_{\rm CM} &= \frac{1}{6A}\sum_{i=1}^{N_v}
\left( x_i + x_{i+1} \right)
\left( x_i y_{i+1} - x_{i+1} y_i \right), \\
\text{y}_{\rm CM} &= \frac{1}{6A}\sum_{i=1}^{N_v}
\left( y_i + y_{i+1} \right)
\left( x_i y_{i+1} - x_{i+1} y_i \right).\end{align}
In the bottom panel of Fig.~\ref{fig:Traj_small}, we find strong qualitative agreement between the droplet trajectories in experiments and simulations, which indicates that the simplified fluid model is appropriate for the experimental studies. 

\begin{figure}[ht]
    \centering
\includegraphics[width=0.95\linewidth]{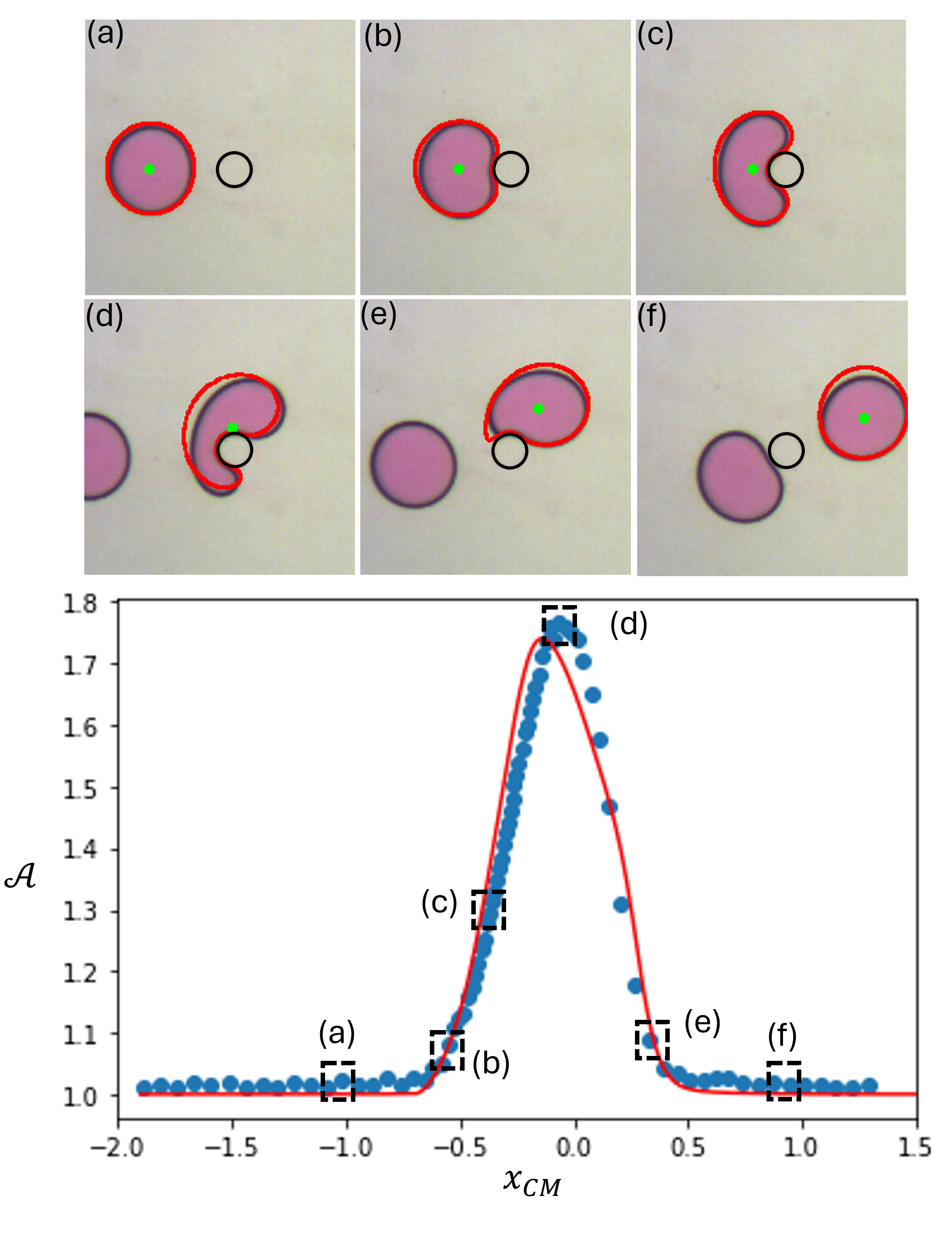}
    \caption{(a)–(f) Snapshots of a large droplet $(D_0/R = 2.5)$ flowing past an obstacle ($R=85\mu$m) in experiments with snapshots from the deformable particle model simulations overlaid in red. The droplet moves from panels (a) to (f) in $\sim 1.05$ seconds. The bottom panel shows the droplet shape parameter $\cal{A}$ plotted versus the $x$-component of the droplet center of mass $x_{\rm CM}$ for the simulations (solid line) and experiments (filled circles).}
    \label{fig:Large_drop_snapshots}
\end{figure}

As the droplet size increases relative to the obstacle size, its motion induces perturbations to the surrounding flow field, leading to distortions of the flow streamlines and higher-order hydrodynamic effects on the droplet dynamics. Resolving these effects would require two-way coupled computational fluid dynamics simulations with moving deformable interfaces. Given that we are focusing on the low Reynolds number regime, viscous dissipation dominates and inertial effects are suppressed, which limits the extent to which droplet-induced flow perturbations affect the droplet motion. 

We can quantify the deformation of the droplet using the dimensionless shape parameter: ${\cal A} = P^2/(4\pi A)$, where $P=\sum_{i=1}^{N_v} l_i$ is the droplet perimeter, ${\cal A}=1$ for a circle, and ${\cal A}>1$ indicates shape-deformation due to interactions of the droplet with the obstacle. In Fig.~\ref{fig:Large_drop_snapshots} (a)-(f), (Supplementary Video 2) we show experimental images of a droplet that is larger than the obstacle with $D_0/R = 2.5$ as it interacts with the obstacle. The bottom panel shows a plot of the shape parameter $\cal{A}$ versus $\rm{x}_{CM}$ of the droplet for both experiments and simulations. Once again, we find qualitative agreement between the simulations and experiments. This level of agreement indicates that, despite its simplifying assumptions, the fluid model in the simulations can capture droplet shape evolution with sufficient accuracy to reliably predict the transition between break up and no-breakup regimes.

\section*{Appendix C: Verifying the Role of Capillary Number in Droplet Breakup} \label{Appendix-C}

\begin{figure}
    \centering
\includegraphics[width=0.95\linewidth]{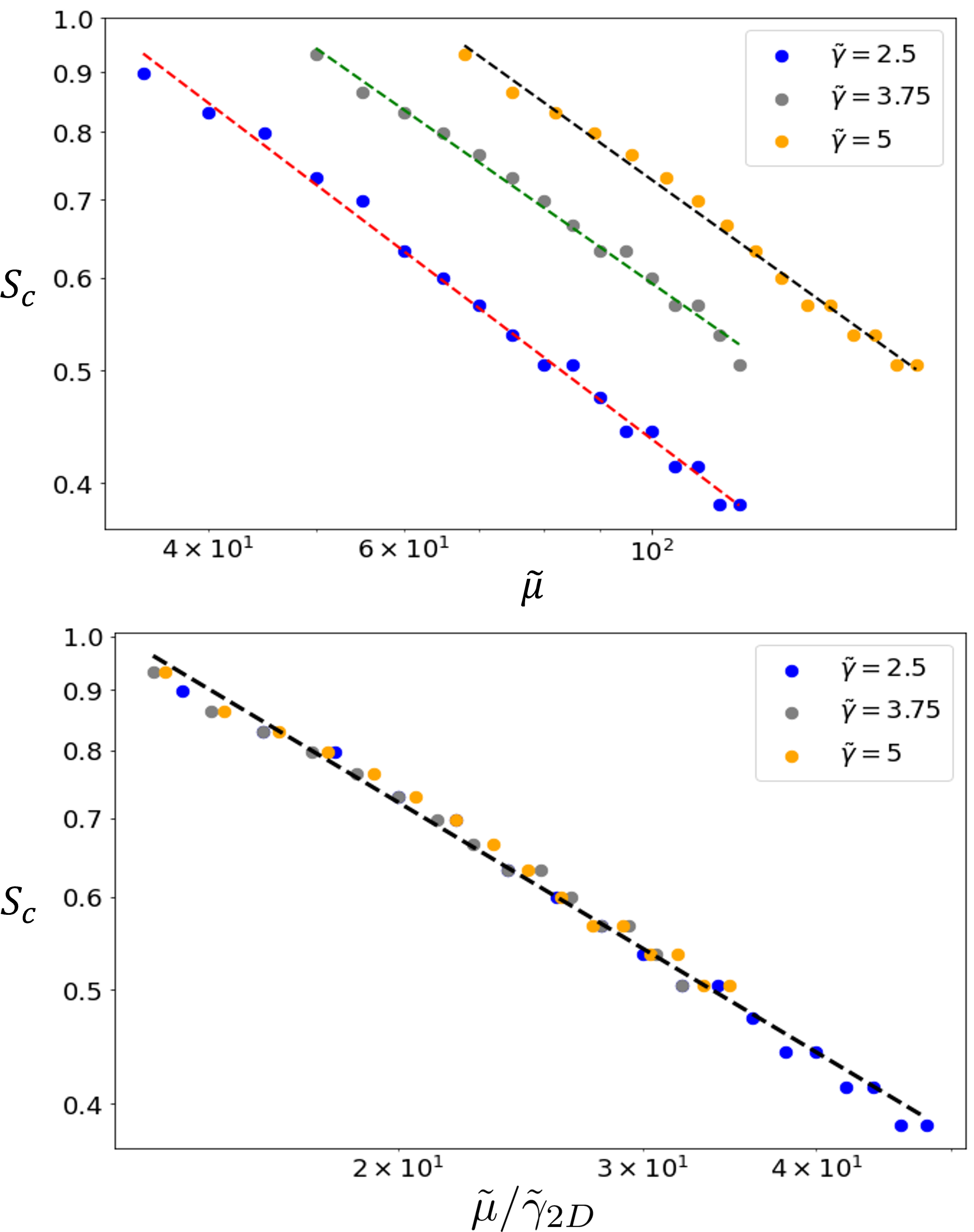}
    \caption{(Top) The boundary in the symmetry parameter $S_c$ that separates break up and non-breakup outcomes for droplet-obstacle collisions as a function of the normalized fluid viscosity ${\widetilde \mu}$ for several values of $\widetilde{\gamma}$. The dashed lines have slope $\beta =-0.72$. (Bottom) The same data in the top panel except the horizontal axis is rescaled as $\widetilde{\mu}/\widetilde{\gamma}$, showing that $S_c$ collapses onto a single curve:  $S_c \sim  \mathrm{Ca}^{\beta}$ (dashed line) with $\beta = -0.72$, which matches the experimental observations of section \ref{separator_regions}.}
    \label{fig:vary_indi_var}
\end{figure}

In this appendix, we describe results from simulations of a single droplet colliding with a single obstacle, while tuning the normalized droplet line tension $\widetilde{\gamma}$ and fluid viscosity ${\widetilde \mu}$ separately. For each $\widetilde{\gamma}$, we vary $S$ and $\widetilde{\mu}$ to determine $S_c(\widetilde{\mu})$ that separates droplet break up from no break up behavior. In the top panel of Fig.~\ref{fig:vary_indi_var}, we show that the separating curve scales as $S_c \sim \widetilde{\mu}^{\beta}$ with $\beta \approx -0.72$ for each value of ${\widetilde \gamma}$. In the bottom panel, we show that all of the $S_c$ curves collapse when the horizontal axis is scaled as ${\widetilde \mu}/{\widetilde \gamma}$. The collapse demonstrates that the combined influence of fluid viscosity and droplet surface tension on droplet break up is governed by the capillary number $\mathrm{Ca} = v \mu / \gamma$.

\section{Acknowledgments}
We thank the laboratories of J.~Burton, C.~Roth, and S.~Urazdhin for use of their equipment in microfluidics fabrication. We thank T.~Brzinski for useful discussions. We acknowledge support from NSF Grant No. CBET-2333224 (D. J. M. and E. R. W.), No. CBET-2333222 (S.S. and C. S. O.), and No. CBET-2333223 (M. D. S.). 



\balance


\bibliography{breakup} 
\bibliographystyle{rsc} 

\end{document}